\documentclass[modern]{aastex62}

\usepackage{graphicx}
\usepackage{amssymb}
\usepackage{amsmath}
\usepackage{epstopdf}
\usepackage{hyperref}
\usepackage{latexsym}
\newcommand{\changes}[1]{\textcolor{black}{#1}}

\received{12 July 2019}
\revised{3 January 2020}
\accepted{5 January 2020}
\submitjournal{ApJ}

\shorttitle{Optimal Filters for Photometric Redshifts}
\shortauthors{Kalmbach et al.}

\begin{document}

\title{Applying Information Theory to Design Optimal Filters for Photometric Redshifts}

\correspondingauthor{J. Bryce Kalmbach}
\email{brycek@uw.edu}

\author[0000-0002-6825-5283]{J. Bryce Kalmbach}
\affil{DIRAC Institute and Department of Astronomy, University of Washington, Seattle, WA 98195, USA}

\author{Jacob T. VanderPlas}
\affil{The eScience Institute, University of Washington, Seattle, WA 98195, USA}

\author{Andrew J. Connolly}
\affil{DIRAC Institute and Department of Astronomy, University of Washington, Seattle, WA 98195, USA}

\begin{abstract}

In this paper we apply ideas from information theory to create a method for the design of optimal filters for photometric redshift estimation. We show the method applied to a series of simple example filters in order to motivate an intuition for how photometric redshift estimators respond to the properties of photometric passbands. We then design a realistic set of six filters covering optical wavelengths that optimize photometric redshifts for $z <= 2.3$ and $i < 25.3$. We create a simulated catalog for these optimal filters and use our filters with a photometric redshift estimation code to show that we can improve the standard deviation of the photometric redshift error by 7.1\% overall and improve outliers 9.9\% over the standard filters proposed for the Large Synoptic Survey Telescope (LSST). We compare features of our optimal filters to LSST and find that the LSST filters incorporate key features for optimal photometric redshift estimation. Finally, we describe how information theory can be applied to a range of optimization problems in astronomy.
\end{abstract}


\section{Introduction} \label{sec:intro}
In a seminal work, \citet{Shannon1948} introduced the concept of information
theory. While originally concerned with the information content of messages sent
along a channel with limited bandwidth and other signal processing problems,
applications of information theory now extend to a multitude of fields
including finance \citep{Ormos+2015} and genomics \citep{Adami2004}. 
Information theoretic concepts are now used in astronomy across a wide range of problems.
For instance, \citet{Weir+1995} worked with decision trees for star/galaxy classification
and used the information entropy to inform the class impurity at each branching.
In \citet{Seehars+2014} the authors utilized information theory to judge the information
gain on parameter posteriors from a series of Cosmic Microwave Background
experiments. They were also able to separate the information gained from improvements in statistical error to that gained 
from new data changing the posterior
distributions.
\citet{Cincotta+1995} proposed the use of Shannon entropy to find
the period of astronomical light curves and \citet{Graham+2013a} extended this to use
conditional entropy. 
\citet{Graham+2013b} showed that the conditional entropy algorithm was the best compared
to a wide variety of other period finding methods including Lomb-Scargle
with regard to period recovery and computation time. Finally, \citet{huijse+2018} used mutual information
to combine light curves measured in different photometric bands and recover periods more effectively than multiband
extensions of Lomb-Scargle and Analysis of Variance periodograms.
In this paper, we apply information theory to a combination of survey design and photometric redshift estimation problems.

Photometric redshift estimation uses multiple observations of extragalactic sources, spread across a
range of filters or passbands, to derive an approximate redshift for a given source \citep{Baum1962, Koo1985, Connolly+1995}. The upcoming Large Synoptic Survey Telescope (LSST) \citep{ivezic+2008} will rely on photometric redshifts for the vast majority of galaxies imaged in the course of its 10-year survey.
The accuracy of these redshift estimates is dependent on the
position of breaks or features within a source spectrum relative to the passbands of the photometric filters. For example, the 4000 \AA\ break transitions out of the LSST
$y$-band at a redshift of $z\sim 1.5$ resulting in an increase in the uncertainty of the estimated redshifts
until the Lyman break enters the $u$-band at $z\sim 2.5$. In principle, the location and shape of a set of filters
could be designed to track specific features within a galaxy spectrum and thereby improve the photometric
redshift (at least over a narrow range of redshifts). This work attempts to find a principled way to define 
the photometric redshift performance of optical filters using information theory and, more specifically,
information gain, and thereby derive a set of filters that are optimal for a specific set of survey objectives. The information gain method
we outline here can be extended in the future to other areas of astronomy where color can be used to classify objects. Here our classes
are photometric redshift bins but could easily be used to classify types of stars instead.

We start in \S \ref{sec:information_theory} with a brief primer on information theory before applying the concept to 3 basic examples in the following sections. \changes{In \S \ref{sec:ig_in_practice} we describe the algorithm and code we developed to calculate information gain for astronomical filters}. In \S \ref{sec:realistic_sample} we apply the technique to design optimal filter sets and in \S \ref{sec:simulated_photoz} we compare photometric redshifts for a simulated catalog using the proposed filter sets versus LSST filters. We discuss our work and future directions for it in Section \ref{sec:discussion} and conclude in Section \ref{sec:conclusion}.

\section{Introduction to Information Theory} \label{sec:information_theory}
\subsection{Entropy} \label{subsec:entropy}
Consider an event $Y$ with a set of $n$ possible outcomes $y_{1}$, $y_{2}$, $y_{3}$, ..., $y_n$ that each occur with probabilities $p_{1}$, $p_{2}$, $p_{3}$, ..., $p_{n}$ and $\sum^n_{i=1} p(y_i) = 1$. How can we measure the amount of choice or uncertainty present in the selection of an outcome? \citet{Shannon1948} concluded that the uncertainty in the observed outcome is given by the entropy ($H$) of this set of possible outcomes where the entropy is defined as:
\begin{equation} \label{eq:entropy}
    H(y) = -\sum_{i} p(y_i) \log_2[p(y_i)]
\end{equation}
Some properties that become apparent from Equation \ref{eq:entropy} are that the maximum entropy occurs when all outcomes are equally probable and that entropy becomes zero when a single probability dominates. Figure 1 shows the entropy when we have two possible choices $A$ and $B$ and how the entropy changes as the probability of getting outcome $A$ changes. When using base 2 in the logarithm then entropy is measured in bits and the entropy represents the average number of binary digits required to encode a set of outcomes from Y.

For instance, imagine we are observing an event that has two possible outcomes that we label $A$ and $B$. If $p(A) = p(B) = 0.5$ then the entropy calculation tells us that the best representation we can derive will encode $H(A) + H(B) = -2 * 0.5 \log_{2}(0.5) = 1$ bit on average. Therefore, simply using $A = 0$ and $B = 1$ when reporting a string of results is an optimal encoding since there is a one-to-one relationship with the length of the encoded information and the number of results. However, if we had a situation where $p(A) > p(B)$ we would have an entropy less than 1. According to information theory then the best encoding scheme could encode the results with less than 1 bit on average. To say this in the reverse way, this means that we can represent a string of results with a number of bits smaller than the length of the results string. Unfortunately, knowing the amount of information in the distribution doesn't tell us how to optimally encode information, but a possible method would be to encode strings of consecutive $A$ results with a single bit. This would mean each bit of information on average would represent more than one result.

\begin{figure}
    \centering
    \includegraphics[width=0.6\textwidth]{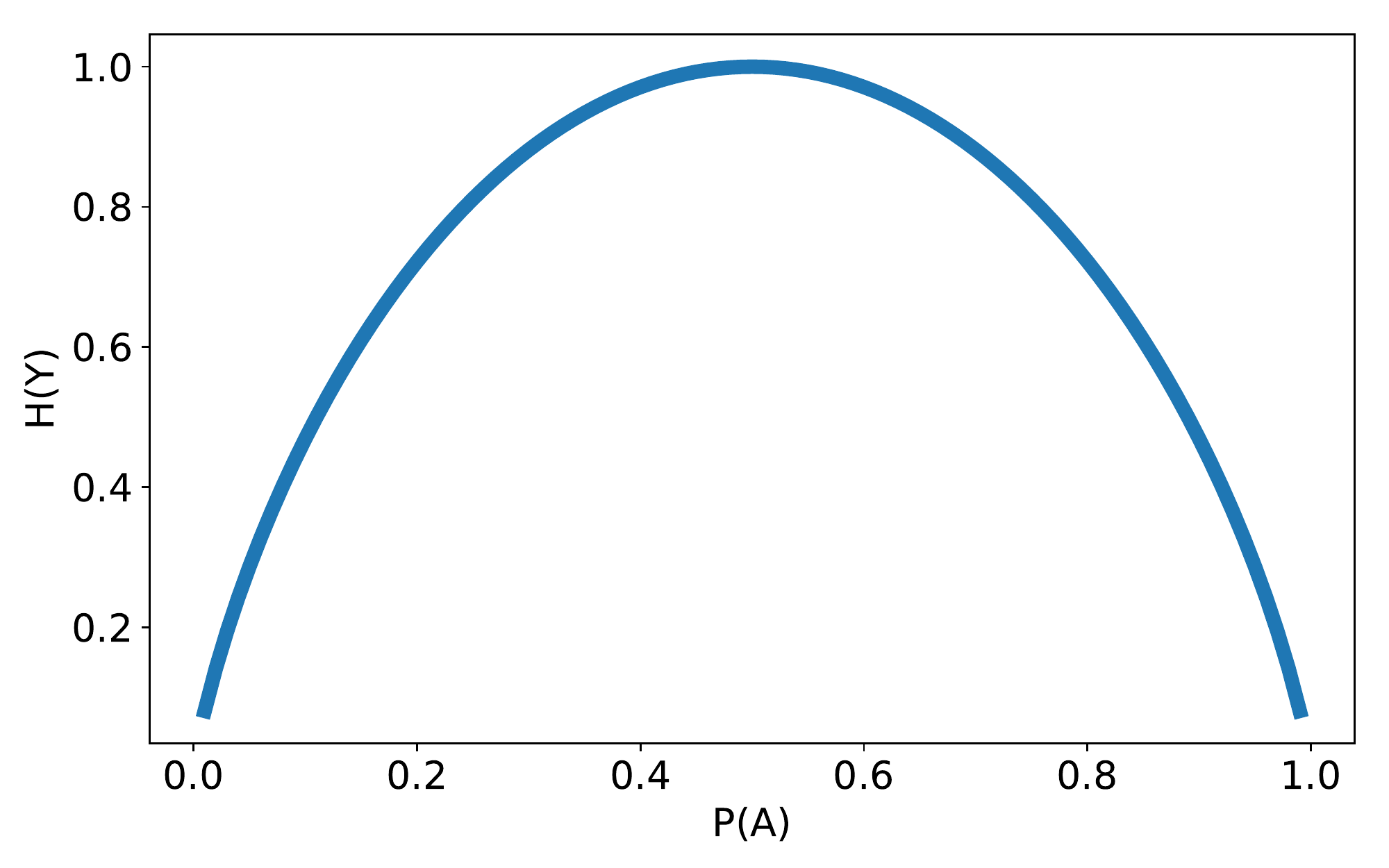}
    \caption{Entropy of a system with two outcomes $A$ and $B$ as the probability of getting outcome $A$ changes.}
    \label{fig:binary_entropy}
\end{figure}

\subsection{Conditional Entropy and Information Gain} \label{subsec:conditional_entropy}
\citet{Lindley1956} was the first to extend information theory to quantify the
information gained from a measurement by measuring how much an experiment
reduced entropy. For instance, imagine a community wants to screen its members for an illness and we know it targets primarily individuals over 40. If we only have a list of the members of the community we can only assign the same probability of illness to all members and can do no better than randomly reaching out to individuals in the population. But if we know the ages of the population we have more information about whom we should target. Using information theory we can actually measure the information gained when the additional information, in this case the ages of the population, is known. To do so we need to know that conditional entropy is the amount of entropy in an observation of $Y$ when the value of $X = x_j$ is a known quantity. It is defined mathematically in a similar way to entropy:
\begin{equation} \label{eq:conditional_entropy}
    H(Y|X=x_j) = -\sum_{i} P(y_i|x_j) \log_2[p(y_i|x_j)]
\end{equation}

In our example, $Y$ is the probability of illness in the overall community and $X$ is the age. If we know the overall distribution of $X$ we can calculate the average conditional entropy for the system:
\begin{equation}
  \label{eq:avg_conditional_entropy}
  H(Y|X) \equiv \sum_{i,j} P(x_j) H(y_i|x_j)
\end{equation}

For the example presented here, we have a different probability for an illness at different ages and this gives us additional information that refines the probability of illness to be more precise for each individual. Therefore, the average conditional entropy will be smaller than the overall entropy since we have less uncertainty in estimates of who might be ill. The actual information gain (IG) can be found by subtracting the average conditional entropy from the original entropy:
\begin{equation} \label{eq:information_gain}
    IG(Y|X) = H(Y) - H(Y|X)
\end{equation}

To put numbers into our example let's give the overall probability of the illness to be 22\%, but for the 60\% of the population under 40 the probability becomes only 10\%, while for the other 40\% of the population it is 40\%. Without information on the ages then we have .76 bits of entropy in our estimates of illness ($.22*\log_{2}(.22) + .78*\log_{2}(.78) = .76$). Adding in the age information gives us an average conditional entropy of $(.6*(.1*\log_{2}(.1) + .9*\log_{2}(.9))) + (.4*(.4*\log_{2}(.4) + .6*\log_{2}(.6))) = .67$ bits. Therefore, the information gain becomes $.76 - .67 = .09$ bits of information gained when we incorporate age information. In the extreme that an illness hit everyone over 40 and nobody under 40 then age information would provide us with a perfect understanding of who had the disease and who didn't. In this case, average conditional entropy would fall to 0 and we would have information gain equal to the total original entropy. This shows that the more information gain we can derive from a measurement of X then the more this measurement reduces our uncertainty in another property Y.

\subsection{Application to Astronomical Observation}
Often in astronomy, we employ a particular observation (be it photometric,
spectroscopic, or other) in order to learn about particular properties
of the object we are observing.  In the formalism expressed above, our
observation (say the magnitude through a particular photometric filter)
is given by $X$, where $X$ represents a continuous distribution of observed
values.  The intended classification of the object (be it 
star/quasar classification, galaxy type, photometric redshift, etc.)\ is
represented by the values $Y$, which may or may not include prior
probabilistic information.  Given a suitably realistic spectroscopic model
of our sample, we can calculate the information gain expected from a
given filter set, and use this quantitative measure to optimize our choice
of filters for the task. In the following sections, we will explore the properties of 
information gain as applied to increasingly more realistic astronomical measurements.

\section{Toy Example 1: Simple Galaxy Classification}
\label{sec:toy_1}

\begin{figure}
 \centering
 \includegraphics[width=.9\textwidth]{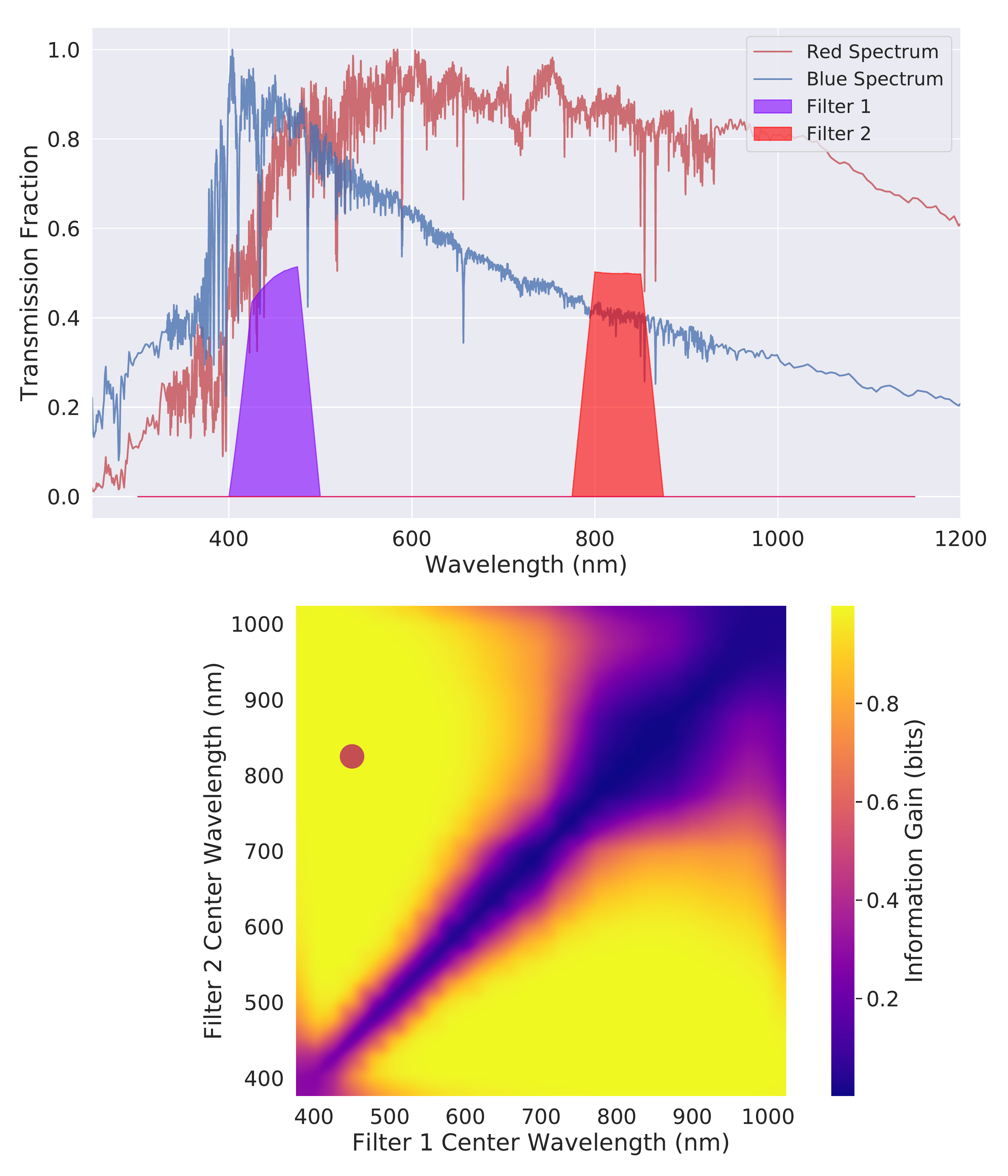} 
 \caption{Top: Optimal filters for differentiating between equally probable red and blue spectra. Bottom: The information gain as a function of central wavelength of each filter. Notice that when the filters are nearly identical the information gain tends towards 0. The maximum information gain filters shown in the top panel are located at the red point with an information gain of over 0.99 bits.
   \label{fig:toy_1} }
\end{figure}

Imagine for the time being that all galaxies have spectral energy distributions (SEDs)
which fall precisely in one of two classes: we'll call them ``red'' and
``blue'' (see Fig.~\ref{fig:toy_1}, upper panel).  
We'll denote this spectral type by the label $Y$,
which can take on the values $Y \in \{y_r, y_b\}$.
Furthermore, imagine that any galaxy has an even chance of being
either red or blue.  Mathematically stated, this means that 
$P(y_r) = P(y_b) = 0.5$.  From Equation~\ref{eq:entropy} we can quickly
compute the entropy $H(Y) = 1$.

Now suppose that an astronomer would like to choose a pair of filters,
the magnitude difference (i.e.\ color)
of which will give maximal discrimination
between the two types of galaxies.  Heuristically, it is clear that
placement of one filter toward the left, and another toward the right
accomplishes this: the difference between the filter magnitudes gives
a positive (red) color for spectrum $y_r$, and a negative (blue)
color for spectrum $y_b$,
leading to an ability to easily distinguish between the two spectra.

This conclusion can be reached in a quantitative fashion
by computing the information gain for
color measurements through the two filters at various locations as
shown in the lower panel of Figure~\ref{fig:toy_1}.  
To construct this surface, we
assume trapezoidal filters (see the upper panel of Figure~\ref{fig:toy_1}) with a total filter response width
of 100 nanometers, containing sloped wings of width 25 nanometers,
and numerically compute magnitudes through each filter. We also include a realistic CCD 
response function that accounts for the curved edges noticeable in the blue end filter 
and the very top of the red filter. We normalize the spectra to $i=22.0$ and include a 
sky background normalized at $i=20.47$. 
Finally, we assume a single LSST visit to calculate 
the magnitudes and the signal to noise ratio (SNR) of each filter. Subtracting the two 
magnitude values for each spectrum gives us the respective colors with a defined set of filters. 
We use the SNR to calculate the expected uncertainty around each color measurement. 
This will give us a Gaussian distribution for the colors of each spectrum in a given filter set. 
The conditional entropy is then calculated by measuring how much the two color distributions overlap.
What we hope to see is that the filter locations that maximize information gain are those that 
move the two colors distributions as far apart as possible.

When we look at Figure~\ref{fig:toy_1} we see exactly that. The top panel shows the 
maximal information gain filters are located with peaks around 450 and 825 nanometers.
The bottom panel is a plot of the information gain as a function of the center wavelength of each
filter and displays that we can almost perfectly distinguish one spectrum from the other since 
our information gain is greater than 0.99 bits out of a possible 1.0. 
The nearly zero information gain along 
the diagonal makes sense since this is where the filters lie on top of one another and produce
the same measured magnitude on average. Information gain is near but not completely zero along this axis
since the width of the error distribution in the color measurement is different for each
spectrum. Therefore, we do have a little bit of information to help label an observation. For
instance, if we make an observation with identical filters and get a color value of 0.02 mags
and this turns out to be a $3\sigma$ measurement for the red spectrum 
but $5\sigma$ for the blue spectrum this provides a small amount of information that increases
the probability of this being a red spectrum measurement.

The sharp rise in information gain from the diagonal in Figure~\ref{fig:toy_1}
is a result of the bright galaxies we used. If the galaxies are fainter
then the increased noise means that the red filter must be further from the peak of the blue spectrum to avoid
the possibility of measuring similar colors for both galaxies. This would present itself as a shallower slope
in the information gain space.

This case showed the basics of the information gain theory with an easy problem. Discriminating 
between two spectra is something we can easily do without resorting to information gain but 
finding the filters that help discriminate between galaxies at different redshifts is 
a more realistic and interesting problem. In the next two sections we move on to two simple examples of optimizing filters for photometric redshift estimation.

\section{Toy Example 2: Measuring the redshift of a sigmoid spectrum}
\label{sec:toy_2}

We can use the same formalism as above to address the question of filter choice
for the determination of photometric redshifts.  In this case the observable
$Y$ is the redshift of the galaxy.  Because $Y$ cannot represent a continuous
distribution within the information gain framework
(note the sums in Equation~\ref{eq:conditional_entropy} above and see Section~\ref{sec:ig_in_practice} below for more information),
we must bin the result.  In practice this is not a problem: using a
sufficiently large number of bins will allow the redshift result to be
recovered to any reasonable accuracy.

For the sake of descriptive simplicity, we'll begin with a toy model
using very simple spectra.
Imagine now that every galaxy in the universe has a spectrum given by
a sigmoid function:

\begin{equation} 
  \label{eq:sigmoid}
  S(\lambda; \lambda_0) = \frac{1}{1 + \exp(\lambda - \lambda_0)}
\end{equation}

This is very close to a step function with $S(\lambda) = 0$ for
$\lambda \ll \lambda_0$, and $S(\lambda) = 1$ for $\lambda \gg \lambda_0$.
With $\lambda_0 = 364.6\ nm$, this shape mimics the balmer-limit break
observed in the spectra of galaxies, from which photometric redshift
determination gains significant leverage.
Imagine furthermore that these galaxies are located at various redshifts,
with a probability distribution given by
\begin{equation}
  \label{eq:prob_z}
  P(z) \propto z^2 \exp[-(z/z_0)^{2}].
\end{equation}
with $z_0$ set so that the median z is 0.6 (typical of ground-based surveys such as DES \citealt{Pogosian+2005}).
If we break the redshift range into 40 bins from $0.0 < z \le 2.$
(giving a bin width $\Delta z = 0.05$) then the information contained
in the redshift of a galaxy can be computed to be $H(Z) \approx 4.4$ 
using Equation~\ref{eq:entropy} and the prior. This can be interpreted
as saying that on average, 4.4 bits of information are needed to specify
the redshift of a particular galaxy in the distribution. Because
there are $40 \approx 2^{5.3}$ bins, one might wonder why a full 5.3 bits
per galaxy would not be needed to specify the redshift: the reason for
this is due to the prior information contained
within the probability distribution (Eqn.~\ref{eq:prob_z}),
which allows a more compact representation of the
data.

\begin{figure}
 \centering
 \includegraphics[width=.9\textwidth]{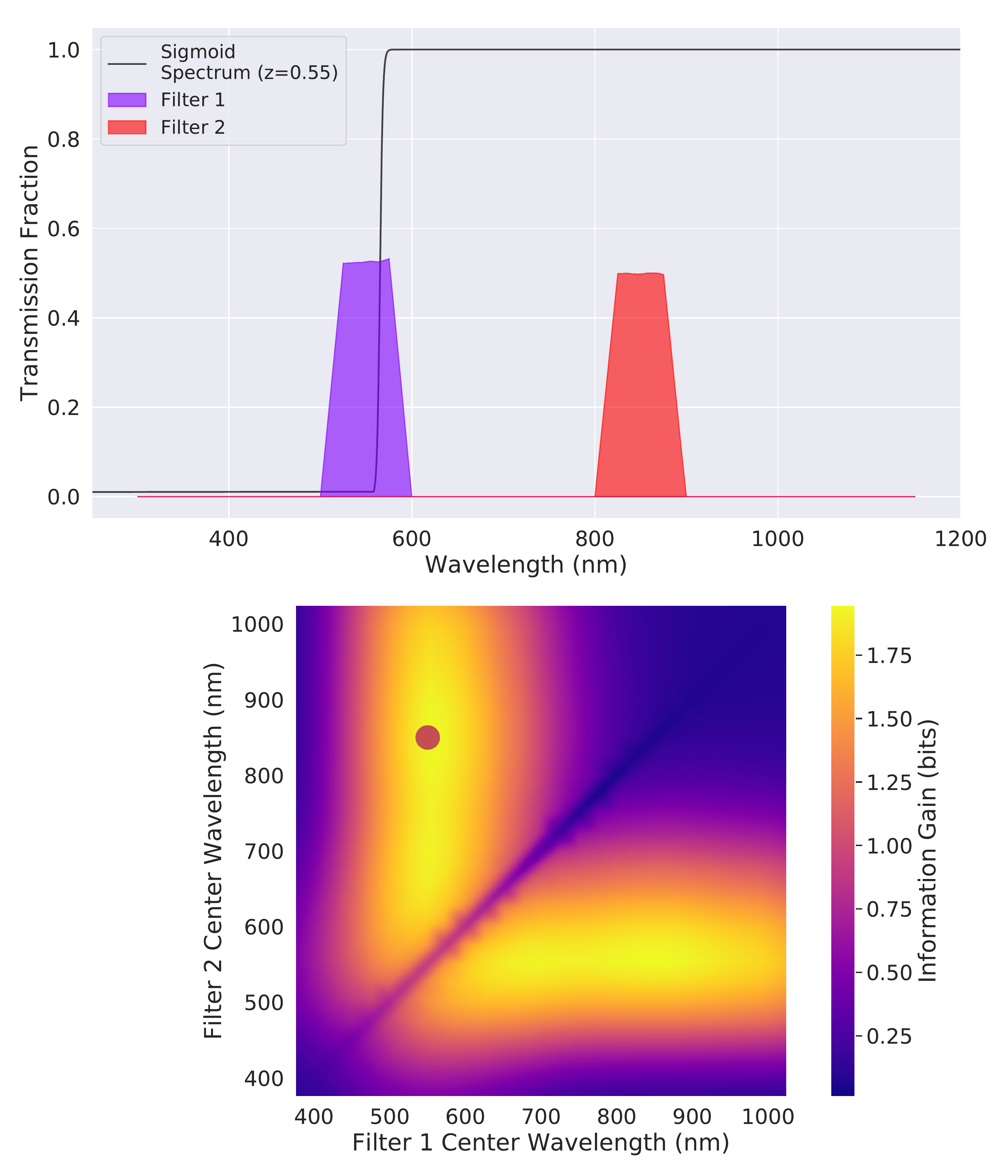} 
 \caption{Top: Optimal filters for differentiating between the sigmoid spectrum at different redshifts up to z = 2. The sigmoid spectrum is shown at a redshift of 0.55 near the peak of the redshift prior function. Bottom: The information gain as a function of central wavelength of each filter. The maximum information gain filters shown in the top panel are located at the red point with an information gain of $\sim 1.95$ bits out of a possible 4.4.
   \label{fig:toy_2} }
\end{figure} 

If we perform a maximization of the information gain
(Eqn.~\ref{eq:information_gain}) using the color observed through
two filters as in Section~\ref{sec:toy_1}, we obtain the information
gain surface shown in the lower
panel of Figure~\ref{fig:toy_2}.  The location of the optimal
filters are much more constrained than in the binary choice example in
Section~\ref{sec:toy_1}.  Because the redshift distribution
peaks near $z = 0.55$, filter combinations where the leftmost filter is centered near
$600.0\ nm$ lead to the greatest information gain,
as seen in the upper panel of Fig.~\ref{fig:toy_2}. 
Since the majority of galaxies are located near the peak at $z = 0.55$ a filter
that can trace the passage of the Balmer break through redshifts near the peak is the most
beneficial.
The broadness of the region of maximal information gain
shows that there is a large
region of the parameter space in which the
filter locations lead to nearly maximal information. As long as one filter is located to measure
the spectral break near the peak of the redshift prior then the other filter can
be shifted left or right over a range of $>200\ nm$ without significantly
reducing the information gain.

Quantitatively, the maximal information gain using two filters
is $\sim 1.95$, out of a total information of roughly 4.40.  That is, in this
simple model, photometric redshifts based on a single color can recover
44\% of the redshift information. Most of the lost information exists because we cannot
easily differentiate between redshifts close to one another when the break is outside
the filters.

\section{Toy Example 3: Measuring the redshift of a single galaxy}
\label{sec:toy_3}

\begin{figure}
 \centering
 \includegraphics[width=\textwidth]{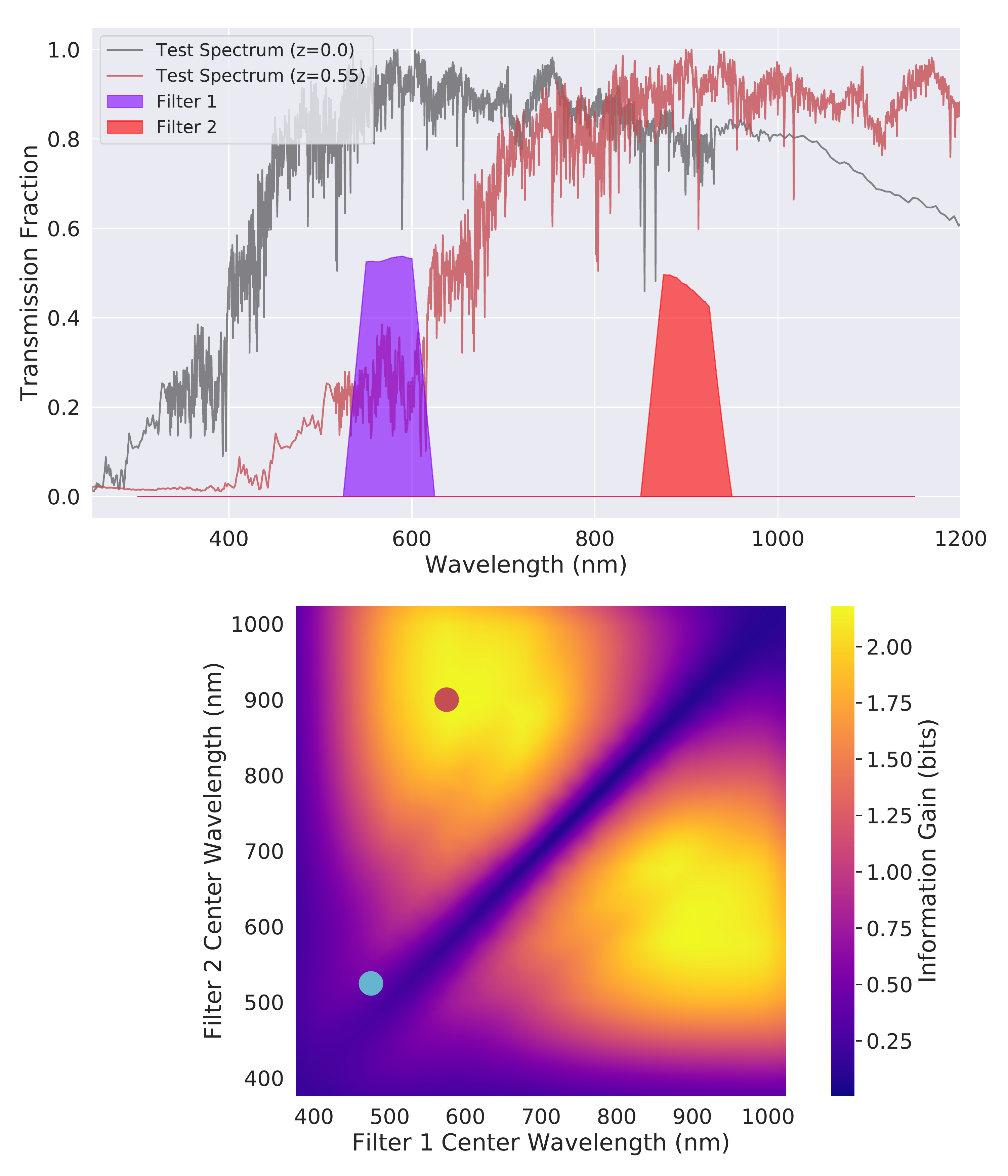} 
 \caption{Top: Optimal filters for differentiating between the red galaxy spectrum at different redshifts up to z = 2.5. Bottom: The information gain as a function of central wavelength of each filter. The maximum information gain filters shown in the top panel are located at the red point with an information gain of $\sim 2.19$ bits out of a possible 4.4. The blue point shows the location of the alternate set of filters used in Figure~\ref{fig:toy_3_colors}.
   \label{fig:toy_3} }
\end{figure} 

\begin{figure}
 \centering
 \includegraphics[width=.9\textwidth]{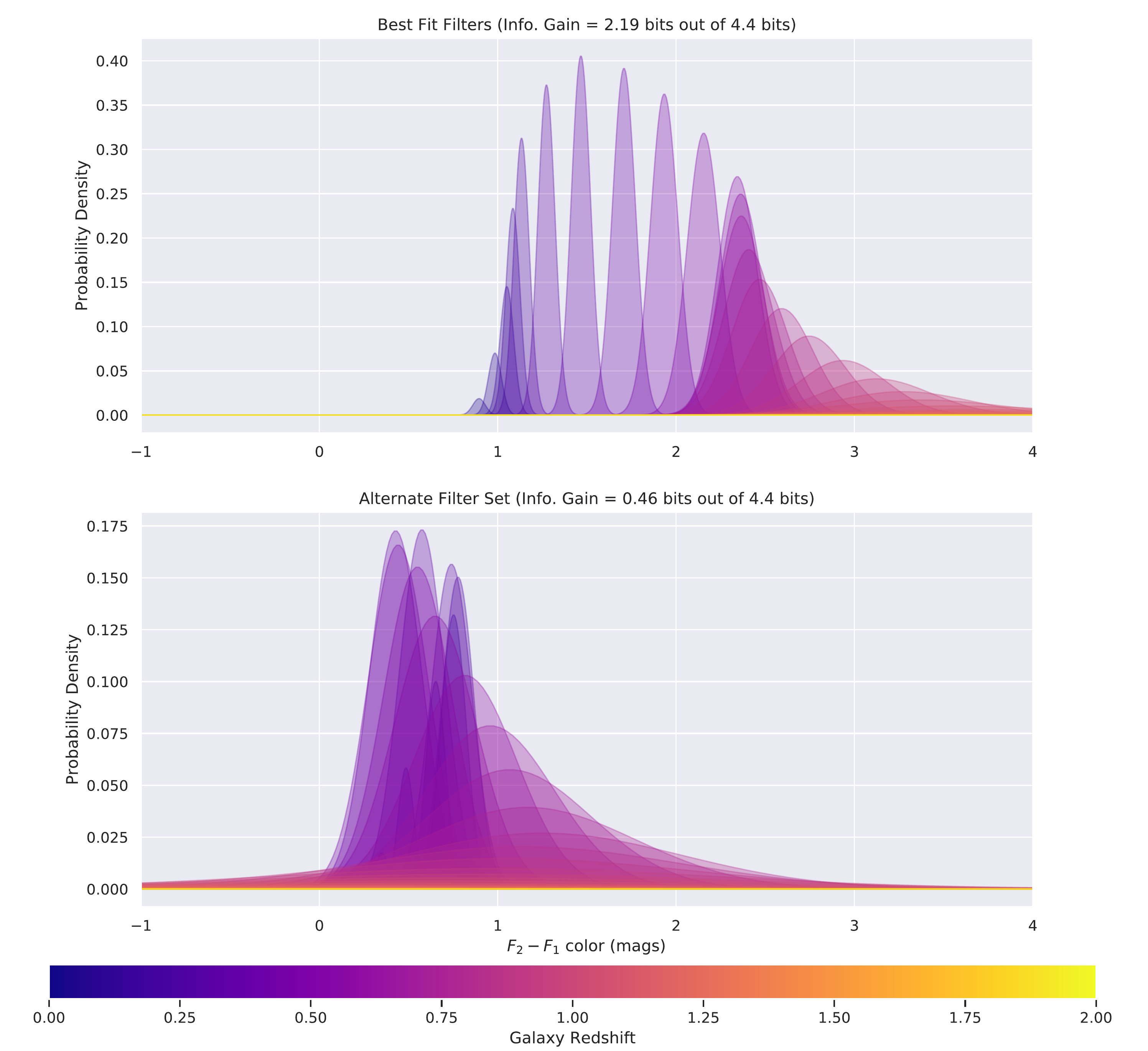}
 \caption{Top: The distribution of colors for the red galaxy spectrum at a series of redshifts using the optimal filter scheme. This figure includes the prior on redshift that more strongly weights intermediate redshifts over low and high redshifts. Notice how redshifts near the peak of the prior ($z\sim0.55$) have the least overlap in their possible color values. Bottom: The distribution of colors using a filter scheme that produces 21\% of the optimal information gain (this corresponds to the blue point in Figure \ref{fig:toy_3}). Here the distributions pile on top of one another and a given color could be the result of the spectrum at a large number of possible redshifts.
   \label{fig:toy_3_colors} }
\end{figure} 

Though the sigmoid spectrum explored in Section~\ref{sec:toy_2} gives some
interesting insight, realistic spectra have many more features in addition
to the Balmer break.  In this section, we explore a similar example using
a single red synthetic galaxy spectrum.

Figure~\ref{fig:toy_3} shows the equivalent of Figure~\ref{fig:toy_2} for
this more realistic spectrum.  The spectrum is that of the red galaxy
from Section \ref{sec:toy_1} with a strong Balmer break around $400\ nm$.  If the
redshift information is coming primarily from this break, we'd expect
the optimal filter locations and associated information gain to be similar
to that seen in the sigmoid spectrum of Section~\ref{sec:toy_2}.

As before, the information gain surface in Figure~\ref{fig:toy_3} shows
a region of low information gain in the places where
the two filters largely overlap.
Also like the previous example, the Balmer break is the main factor that
determines the locations of the optimal filters. The bluer filter
is once again located in the range where the break is passing through the filter
when the spectrum is redshifted to the peak redshift of the prior distribution.
The maximal information gain in this
case is slightly higher to what we saw previously: $\sim 2.19$ out of 4.40.
This increase is mainly due to the broadness of the Balmer jump (about $100\ nm$ wide
here compared to a negligible width previously) which allows a wider range of redshifts
to benefit from the change in magnitude of the blue filter as the break passes through
it.

To see exactly what is the difference that leads to better information gain
in one set of filters versus another we explain the results shown in
Figure~\ref{fig:toy_3_colors}. Here the top panel shows the probability
distributions of observed colors of the spectrum at
each redshift using the optimal filters from
Figure~\ref{fig:toy_3} weighted by the prior probability at that redshift. These probability
distributions are centered at the mean color for the template at a given redshift. Since the 
redshift is binned and calculated in steps of 0.05
we see a discrete set of color distributions.
The width of each color distribution is a result of 
photometric uncertainties and is affected by the design of the filters and survey. 
In the bottom panel we see the same distributions for colors derived
using a set of filters that only produced 0.46 bits of information gain and are
marked in the lower panel of Figure~\ref{fig:toy_3} by the blue dot. Notice how
much more overlap there is in the distributions for colors at each
redshift in the bottom panel. On top where we have
higher information gain we can be much more confident that a galaxy
measured with a certain color will have a given redshift especially in the redshifts around the
peak of the prior distribution at $z \sim 0.55$. 

The simple examples shown here help connect the information theory presented to practical results
in astronomical terms. However, to fully apply information theory to larger template sets and
higher numbers of filters we need to further develop our mathematical approach and build the computational
tools that will allow us to perform larger experiments.

\section{Calculating Information Gain in Practice} \label{sec:ig_in_practice}
To calculate the information gain in more complicated scenarios we needed to develop code that could quickly calculate information gain (IG) based upon multiple colors and redshifts of multiple SEDs. For this purpose we created a python code called \textit{SIGgi} \citep{Kalmbach2019} (where SIG stands for Spectral Information Gain). In practice we calculate IG starting from calculating the average conditional entropy $H(Y|X)$ where we rewrite it by combining Equations \ref{eq:conditional_entropy} and \ref{eq:avg_conditional_entropy} along with the identity $P(x_i, y_i) = P(y_i|x_i) P(x_i)$ to get:
\begin{equation} \label{eq:expanded_conditional_entropy}
    H(Y|X) = -\sum_{i,j} p(x_j, y_i) \log_2[\frac{p(x_j, y_i)}{p(x_j)}]
\end{equation}
But in our case $X$ is the vector of colors of the SED and is continuous. Therefore, we allow continuous observations by using the Kullback-Leibler (KL) divergence \citep{kullback+1951}:
\begin{equation}
    D_{KL}(P||Q) = \sum_k p(y_k) \log_2(p(y_k)/q(y_k))
\end{equation}
If $p(y)$ and $q(y)$ are continuous probability distributions and normalized to 1 across the entirety of $y$ then the KL divergence is:
\begin{equation}
    D_{KL}(p||q) = \int p(y) \log_2[p(y)/q(y)]\ dy
\end{equation}
Now we can see that $H(Y|X)$ can be written in terms of the KL divergence as:
\begin{equation}
    H(Y|X) = - D_{KL}(p(y_i, x) || p(x))
\end{equation}
where $y$ remains a discrete variable and thus requires that we continue to bin redshift, but now $x$ is expressed as a continuous observable. Finally we combine this with Equation \ref{eq:expanded_conditional_entropy} to get:
\begin{equation} \label{eq:continuous_coditional_entropy}
    H(Y|X) = -\sum_i \int p(y_i, x) \log_2 [\frac{p(y_i, x)}{p(x)}]\ dx
\end{equation}
where $i$ is a particular redshift bin.

So, to compute the conditional entropy we must be able to determine the
joint probability $P(y_i, x) = P(y_i) P(x|y_i)$, where each
$y_i$ represents
a discrete unknown property (e.g.\ binned redshift), and $x$ is a continuous
observable (e.g.\ photometric colors).
$P(y_i)$ is simply the prior distribution of the unknown property,
and $P(x|y_i)$ expresses the distribution of observables for a particular
input.
We have a model to predict this conditional distribution $P(x|y_i)$
of an observation $x$ given a value $y_i$ (for example, we can compute the
colors of a galaxy given its redshift).  In the case of a single spectrum
this distribution $P(x|y_i)$ is assumed to be normally distributed about a 
single value.  The width of the Gaussian in each dimension will
depend on the photometric uncertainty of the measurements in the filters for that color.
In the case of multiple spectra the color distribution for a single redshift
will be the sum of the normal distributions for each individual spectrum
at that redshift.

\begin{figure}
 \centering
 \includegraphics[width=\textwidth]{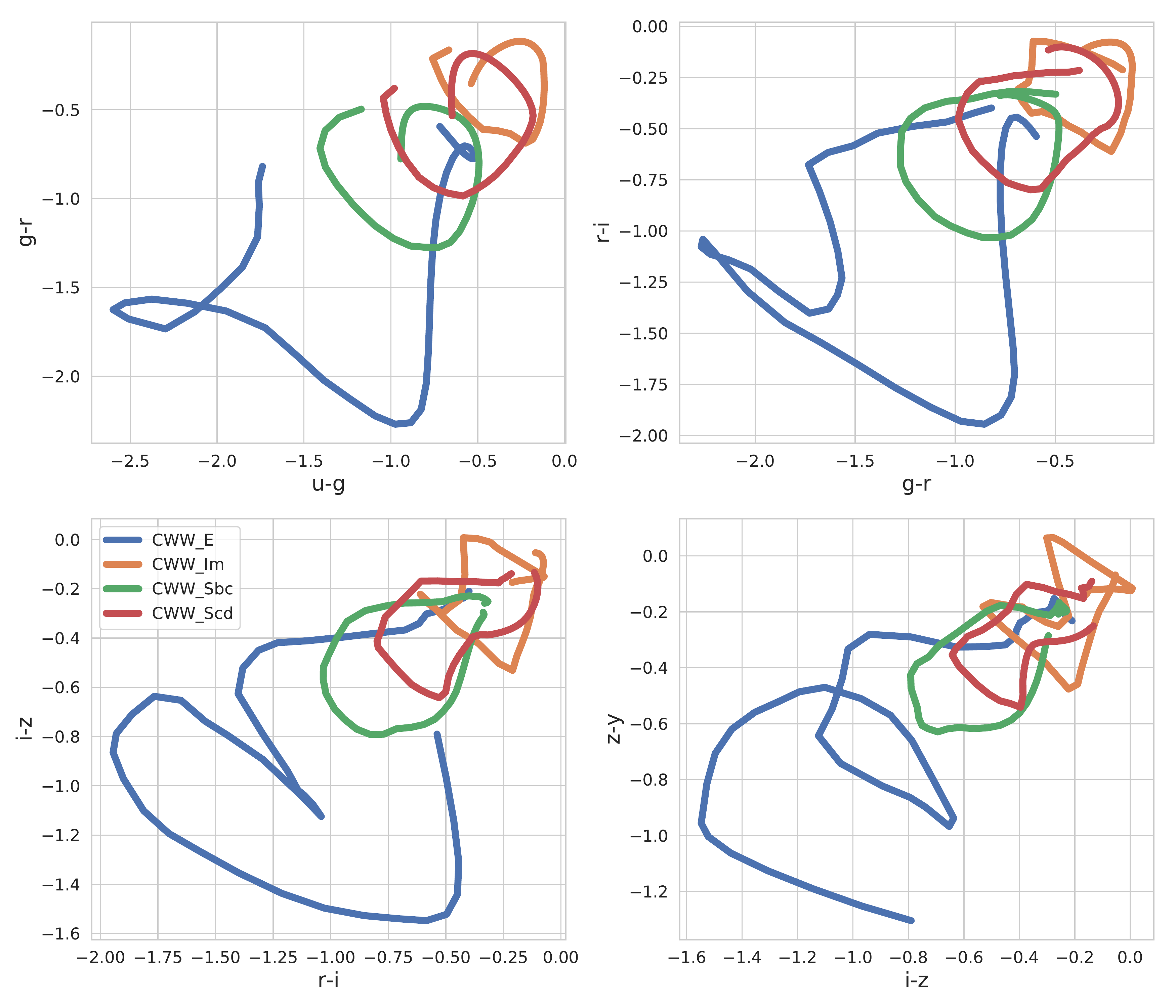}
 \caption{Color-color plot of the 4 \citet{Coleman+1980} templates in the LSST filters.
   \label{fig:lsst_color_color} }
\end{figure}

Because the calculation of conditional entropy via
Equation~\ref{eq:continuous_coditional_entropy} involves an integral over a
potentially high-dimensional space with very fine resolution, a
straightforward numerical integration based on a grid of values
becomes too costly to use in practice.  For example, a single color for a 
collection of galaxy spectra
in LSST filters may spread over up to two and a half magnitudes (see Figure~\ref{fig:lsst_color_color}).
To assure sufficient sampling
of the distribution of colors, this requires on order 100 grid divisions per dimension,
which leads to on order $10^{10}$ grid points in five dimensions for
a naive implementation.  In practice, even this resolution can produce
artifacts due to insufficient sampling of the distributions.

However, the calculation of this integral can be done by probabilistically sampling from the color
distributions. We start from Equation \ref{eq:continuous_coditional_entropy} in combination with the approximation $\int p(x) q(x)\ dx = \lim_{N \xrightarrow{} \inf} \frac{1}{N} \sum_{x_i \sim p(x)} q(x_i)$ and calculate the following in our code:
\begin{equation} \label{eq:sampling_conditional_entropy}
    H(Y|X) = -\frac{1}{N} \sum_i \sum_{x_j \sim p(y_i, x_j)}^{n_{i}} \log_2[\frac{p(y_i, x_j)}{p(x_j)}]
\end{equation}
To evaluate Equation~\ref{eq:sampling_conditional_entropy}, 
we draw $N=10^6$ points from the prior distribution for 
redshift $p(y_i)$. This gives us $n_i$ points that fall
into each redshift bin that we then use to calculate the inner
sum for that bin. For each point in the redshift bin we
randomly pick an SED with a uniform probability (a simplification
we discuss modifying in future work in Section \ref{sec:discussion}).
We then draw a vector of colors from the multivariate Gaussian distribution
that models the observed photometric color and uncertainties for the redshifted SED.
We save all these color points 
so that we have a representation of the complete color space
for that redshift based upon the available galaxy SEDs. 

To calculate the sum over the logarithm we need to find the values for
$p(y_i, x_j)$ and $p(x_j)$
where $x_j$ is a point in color space. The value for $p(x_j)$ will be the sum
of values measured at $x_j$ from the multivariate Gaussians that are 
the probability density functions for the colors for each SED at each redshift.
To get $p(y_i, x_j)$ this calculation includes
only the redshifted SEDs at the specific redshift $y_i$. 
We use this technique to sum over the
points in each redshift bin and then sum over the values for
each redshift bin before normalizing by $\frac{1}{N}$ to get
a final answer for $H(Y|X)$. This value for conditional entropy
is then subtracted from the full entropy to get the information gain.

\section{A Realistic Sample}
\label{sec:realistic_sample}
The real world is not nearly as clean as the simple situations discussed above.
Rather than observing galaxies of a single spectral type, we observe
many different galaxies with different intrinsic spectral characteristics
at many different redshifts. Rather than having
a single color associated with each redshift, we have a broad distribution
of colors associated with each redshift.

\begin{figure}
    \centering
    \includegraphics[width=\textwidth]{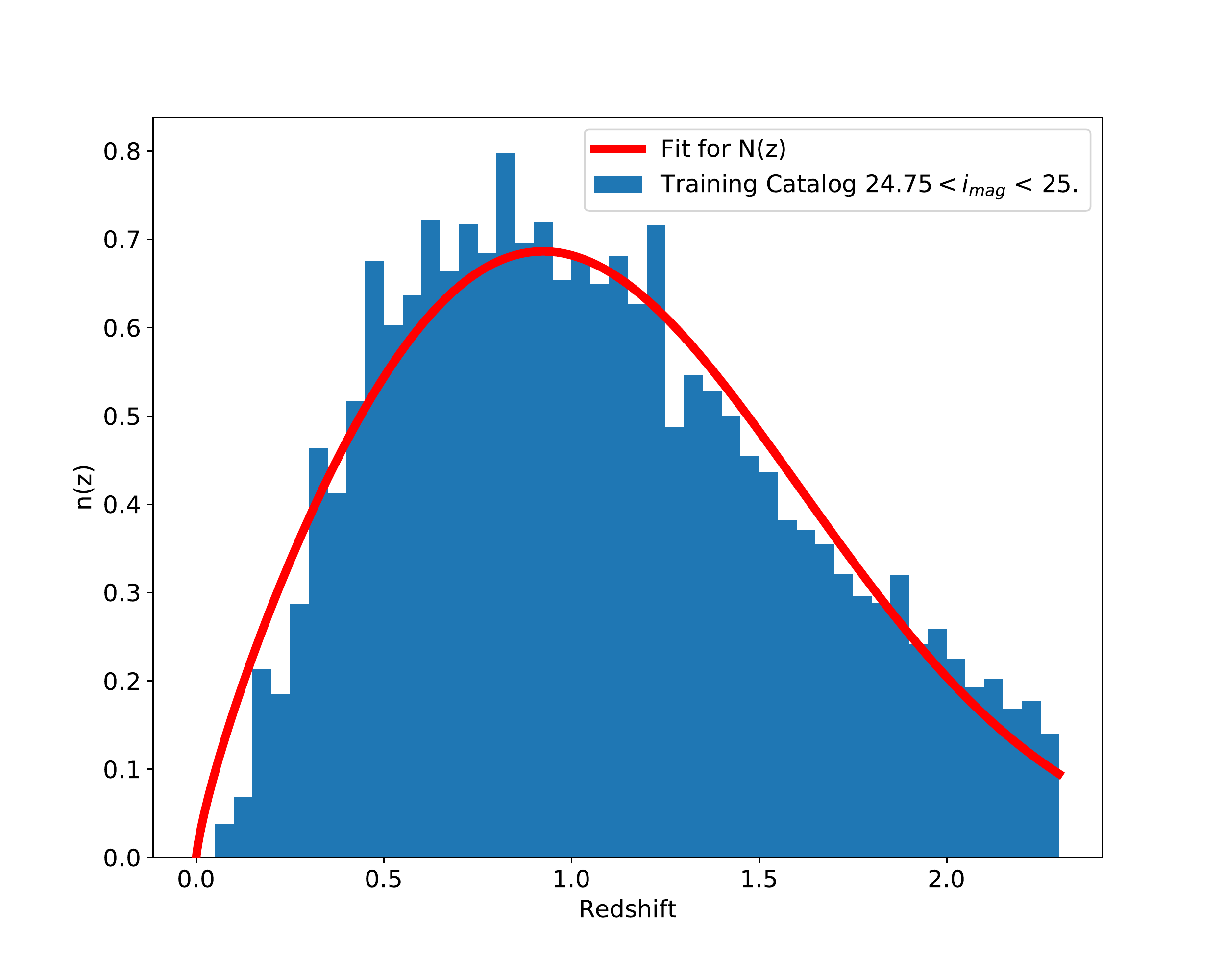}
    \caption{Redshift prior derived from training catalog.}
    \label{fig:catsim_prior}
\end{figure}

To study this, we need a representative sample of spectra which evenly
samples the expected space of observations. Since we are interested
in the estimation of photometric redshifts we use
template sets from \citep{Coleman+1980} (CWW) and \citep{Calzetti+1994}
(Kinney-Calzetti Atlas) supplemented by \citet{Arnouts+1999} at UV and IR
wavelengths with the GISSEL code \citep{Bruzual+1993, Bruzual+2003}. 
The colors and photometric uncertainties for the Gaussian distributions in color space are 
calculated based upon normalizing all the SEDs to $i=25.3$ and using a sky
background set at $i=20.47$ with an LSST-like telescope over a 10 year LSST-like survey.
The sky background is modeled
with a sky SED provided in the LSST Sims throughputs package \citep{Connolly+2014}. 
We choose $i=25.3$ for our normalization since this matches the faint 
limit on the definition of LSST ``gold sample" galaxies for photometric 
redshifts \citep{LSSTScienceBook}. We also use a prior function on redshift
that we derive from the photo-z training catalog we describe in 
Section \ref{subsec:sim_catalog}. This prior came from fitting a function
of a similar form to Equation \ref{eq:prob_z} to the 39,952 training 
set galaxies with $24.75 < i_{mag} < 25.$ and is designed to approximate the
galaxies expected near the faint normalization of the SEDs. The histogram
of the galaxies and the prior are shown in Figure \ref{fig:catsim_prior}. 
In our tests, we bin the redshift every 0.05 between $0.0 \leq z \leq 2.3$ 
giving us 46 total bins. We set the limit at 2.3 because the CWW-Kinney 
templates blue limit starts to pass into the bluest wavelengths of
our filters at this redshift.

Finally, instead of sampling along a grid of set widths and centers as we did in the example problems we use scikit-optimize \citep{scikit-opt} to optimize the locations and shapes of our filters. Scikit-optimize is an open source python-based Bayesian optimization package designed to optimize complex spaces such as the high-dimensional information gain space in our problem. We use the Gaussian Process based estimator provided in the code to model the output space and choose locations for optimization. In each run, we initialize the space with 10 points before running the optimization and allow the optimization to run in parallel, updating after running a set number of points independently each time.

\subsection{Adding a new filter to LSST} \label{subsec:lsst_plus_one}

In our first experiment we used the LSST filters as a set of fixed filters and 
wanted to find the optimal additional filter in the optical range that would
benefit photometric redshifts the most. For our simulation we gave this filter an equal number of visits 
as proposed for the LSST $y$ filter and kept the same number of visits for the other
filters in effect extending the baseline LSST survey shown in Table \ref{tab:lsst_visit_table}. We allowed the four corners of a trapezoidal filter to move independently in the wavelength range between 300 and 1100 $nm$. This gave us an optimization with four degrees of freedom that allowed the location and width of the filter to vary as well as the slope of the wings of the filter on each side.

\begin{table}[]
\centering
\caption{Number of visits to a field in LSST survey for each filter}
\label{tab:lsst_visit_table}
\begin{tabular}{l|c|c|c|c|c|c}
\hline
Filter       & \textbf{u} & \textbf{g} & \textbf{r} & \textbf{i} & \textbf{z} & \textbf{y} \\ \hline
\# of visits & 56         & 80         & 184        & 184        & 160        & 160        \\ \hline
\end{tabular}
\end{table}

\begin{figure}
 \centering
 \includegraphics[width=.9\textwidth]{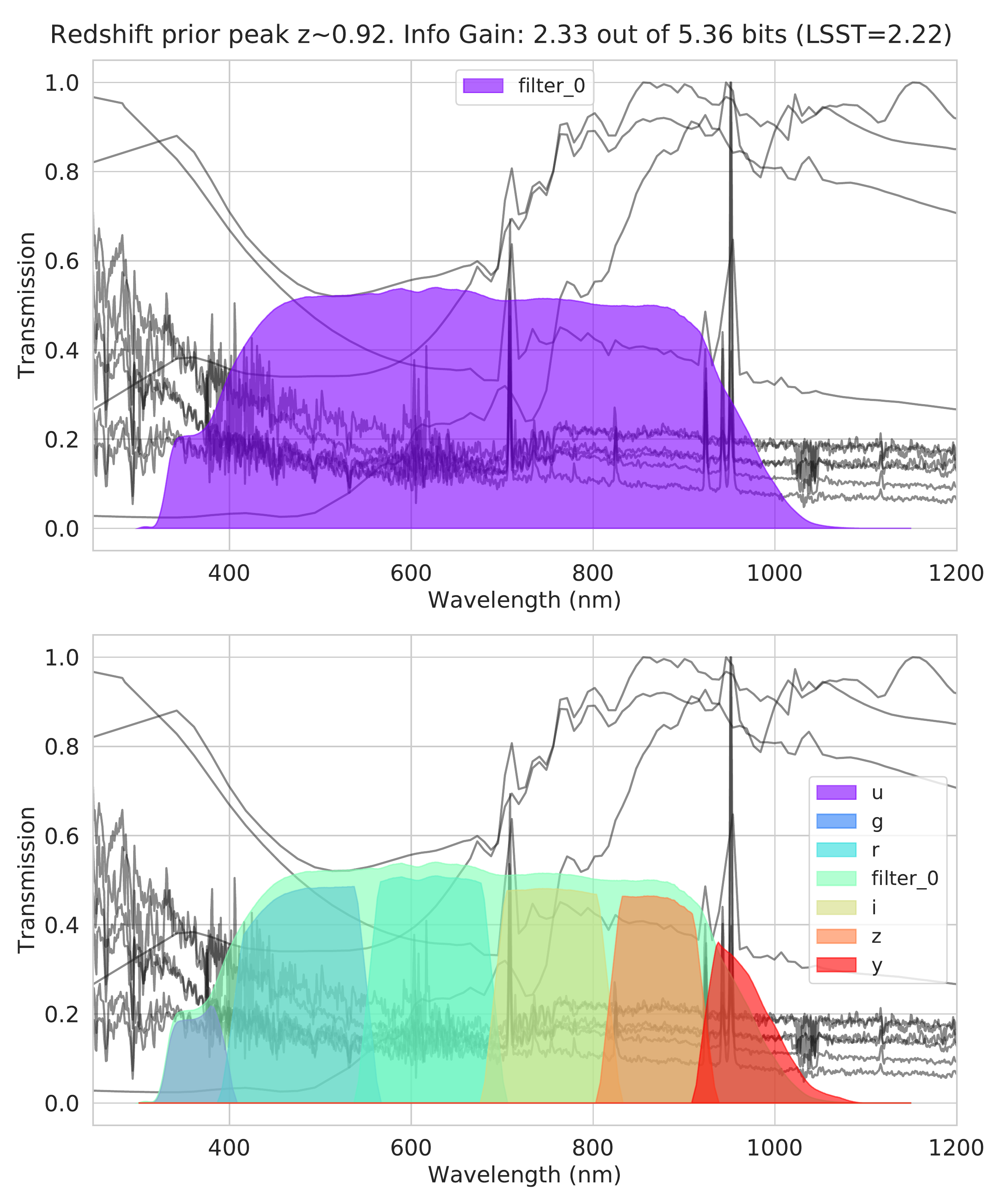} 
 \caption{Top: The best additional filter added to LSST filters is a wide filter overlapping all the original LSST filters when the template flux normalized to LSST $i = 25.3$. The CWW-Kinney templates are shown in the background redshifted to the peak of the prior distribution ($z \sim 0.92$). Bottom: The additional filter with the LSST filters provided for comparison.
   \label{fig:plus_one_filter_wide} }
\end{figure}

The resulting filter is shown in Figure~\ref{fig:plus_one_filter_wide} with and without the
accompanying LSST filters. The best filter is a large filter with
wide wings at the blue and red ends. This filter raises the information gain only slightly
from 2.22 bits for the LSST filters alone to 2.33 bits. 

This wide filter is centered around the Balmer break at the peak redshift of the 
prior distribution at $z \sim 0.92$. This is obvious in the top panel of 
Figure~\ref{fig:plus_one_filter_wide}
where the test SEDs are shown redshifted to this peak value. This seems to confirm
what we saw in the examples of the Sections \ref{sec:toy_2} and \ref{sec:toy_3}.
Another thing to notice is that the additional information gain provided by
a seventh filter to the LSST in the optical range is only a 5\% improvement.
This indicates that it is difficult to improve the LSST filters by adding wide
filters in the optical range.

\changes{To understand what is driving the design of the seventh filter we ran 
the same experiment but normalized our templates to LSST $i=23.0$. In this case, 
shown in Figure 9, the filter narrows to focus on the region around the Balmer 
break. With brighter galaxies we are able to get a higher signal-to-noise 
measurement of the wavelengths around the Balmer break at the peak of the prior 
with a narrower filter. This indicates that the original, wider seventh filter 
is a result of trying to maximize the signal-to-noise of the faint galaxy 
populations over improving the redshift sensitivity through the introduction of 
a narrower filter.}

\begin{figure}
 \centering
 \includegraphics[width=.9\textwidth]{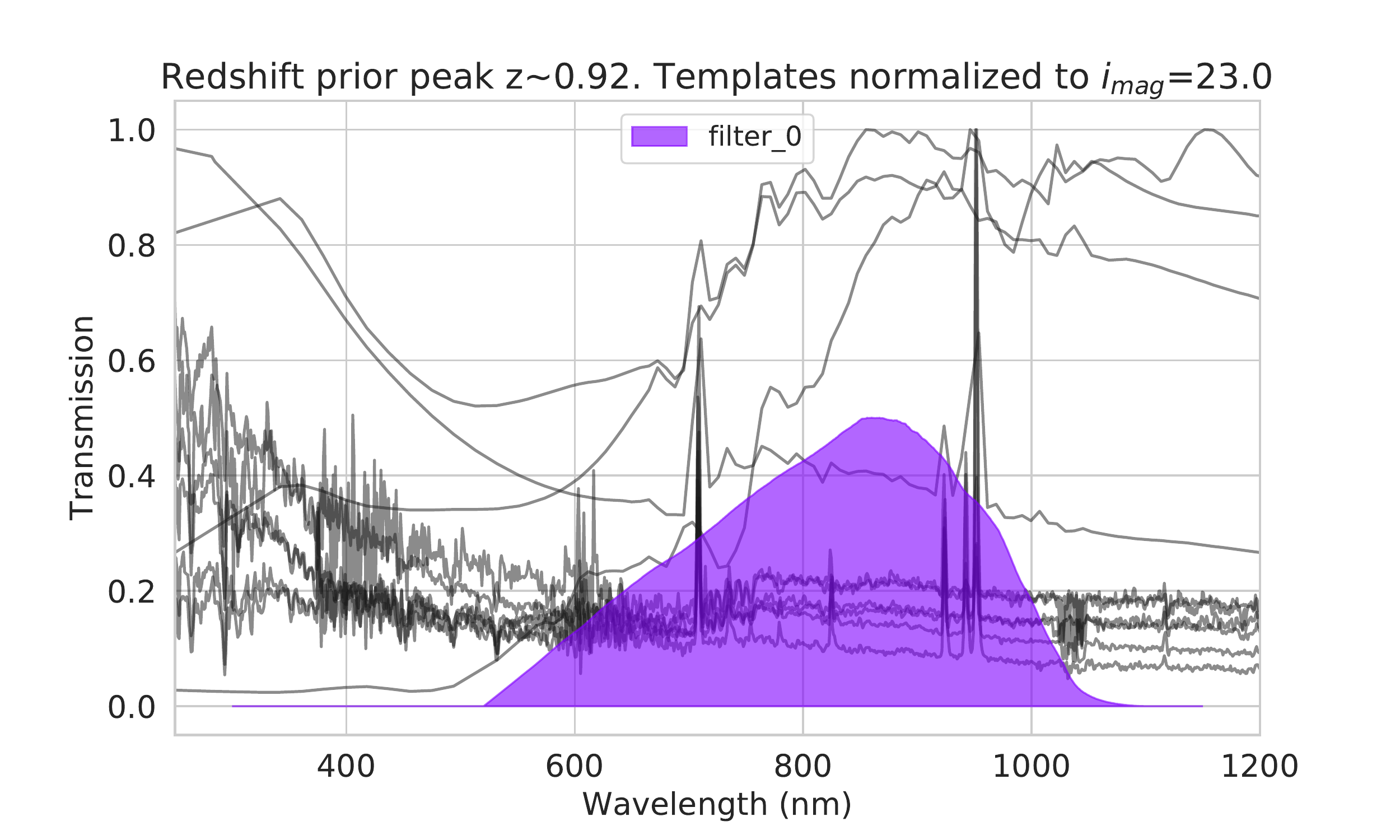} 
 \caption{The best additional filter added to LSST filters when the template flux normalized to LSST $i = 23.0$. The filter narrows to focus on the region of the Balmer break.
   \label{fig:plus_one_filter_bright} }
\end{figure}

\subsubsection{Effect of changing prior}

\begin{figure}
 \centering
 \includegraphics[width=.9\textwidth]{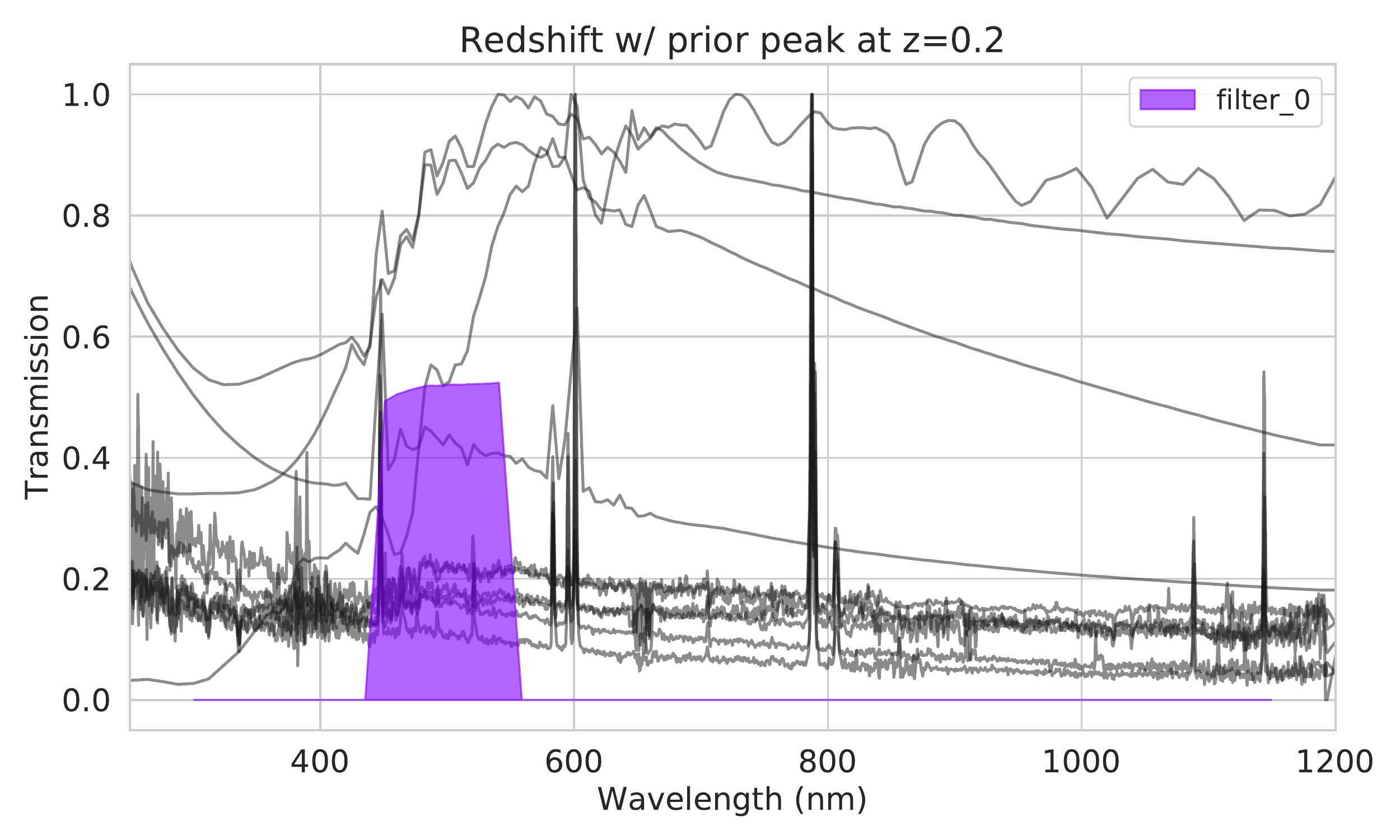} 
 \caption{The best additional filter when using a redshift prior distribution that peaks at
$z \sim 0.2$. The optimal filter is shifted further towards the blue end of the optical range
to get information from the Balmer break around the peak redshift of the redshift prior. The SED
templates are shown in the background redshifted to $z = 0.2$.
   \label{fig:plus_one_alt_prior} }
\end{figure}

To verify that the Balmer break is the primary source of information we reran the optimization
with a different prior to see how the location of the seventh filter changed. We created a toy 
prior that peaks at $z \sim 0.2$. 
The outcome is shown in Figure~\ref{fig:plus_one_alt_prior} and confirms the shift in the
filter location toward the location of the Balmer break at the new peak of the redshift prior.
Thus, we observe that filters will constrain redshift the best if they can optimally
constrain the location of the Balmer break as it moves across the optical wavelength range.

\subsection{Six filter survey: Properties of optimal filter sets for photometric redshifts} \label{subsec:six_new_filters}

The locations and shapes of photometric filters affect the colors observed for stars and galaxies. Photometric redshifts rely upon the design of filter systems that will pick up the spectral features for galaxies in the relevant redshift range of a survey. The colors produced by a photometric system are also important for estimating stellar properties \citep{Lenz+1998} and quasar selection \citep{Peters+2015}. Here we investigate optimal shapes and locations for photometric redshifts but plan to extend this evaluation to other astronomical problems that require colors in future work.

\begin{figure}
 \centering
 \includegraphics[width=\textwidth]{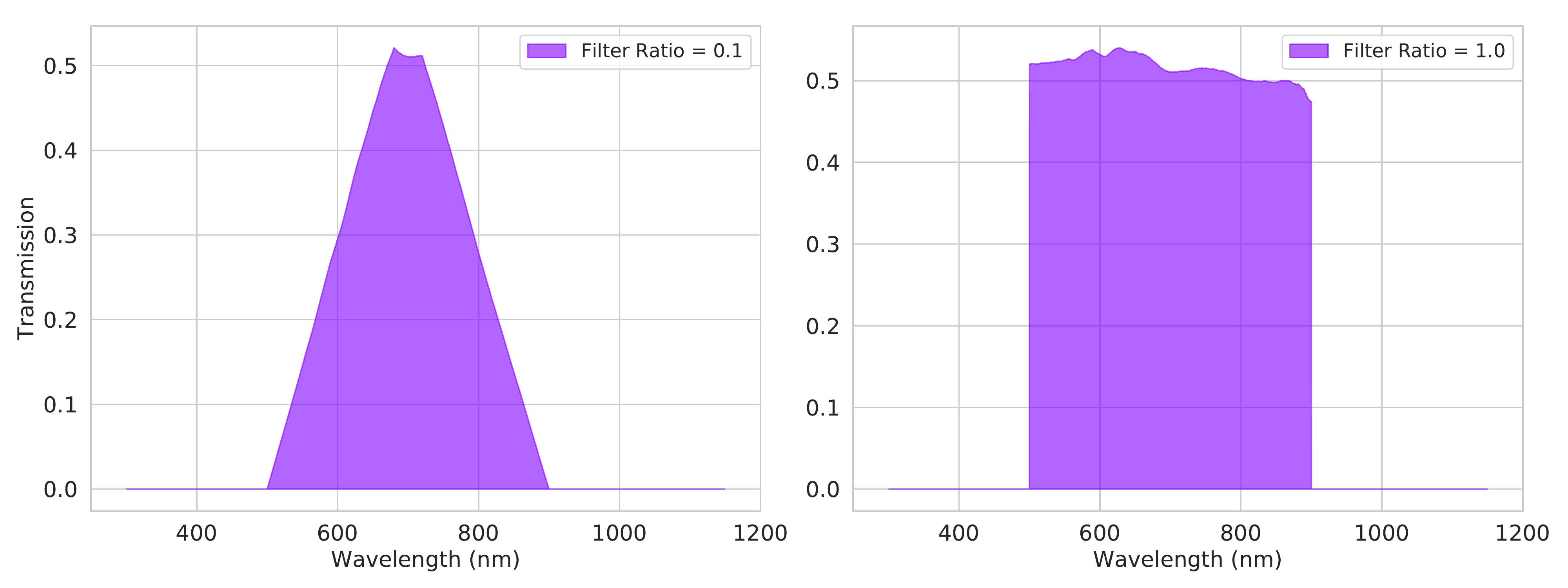} 
 \caption{A comparison of the allowable filter shapes. Left: A filter with a ratio of top width to bottom width of 0.1. Filters with lower ratios are more triangular. Right: A filter with a top-to-bottom width ratio of 1.0. Filters with higher ratios are more rectangular or "top hat" like. 
   \label{fig:filt_shape_compare} }
\end{figure}

\begin{figure}
 \centering
 \includegraphics[width=\textwidth]{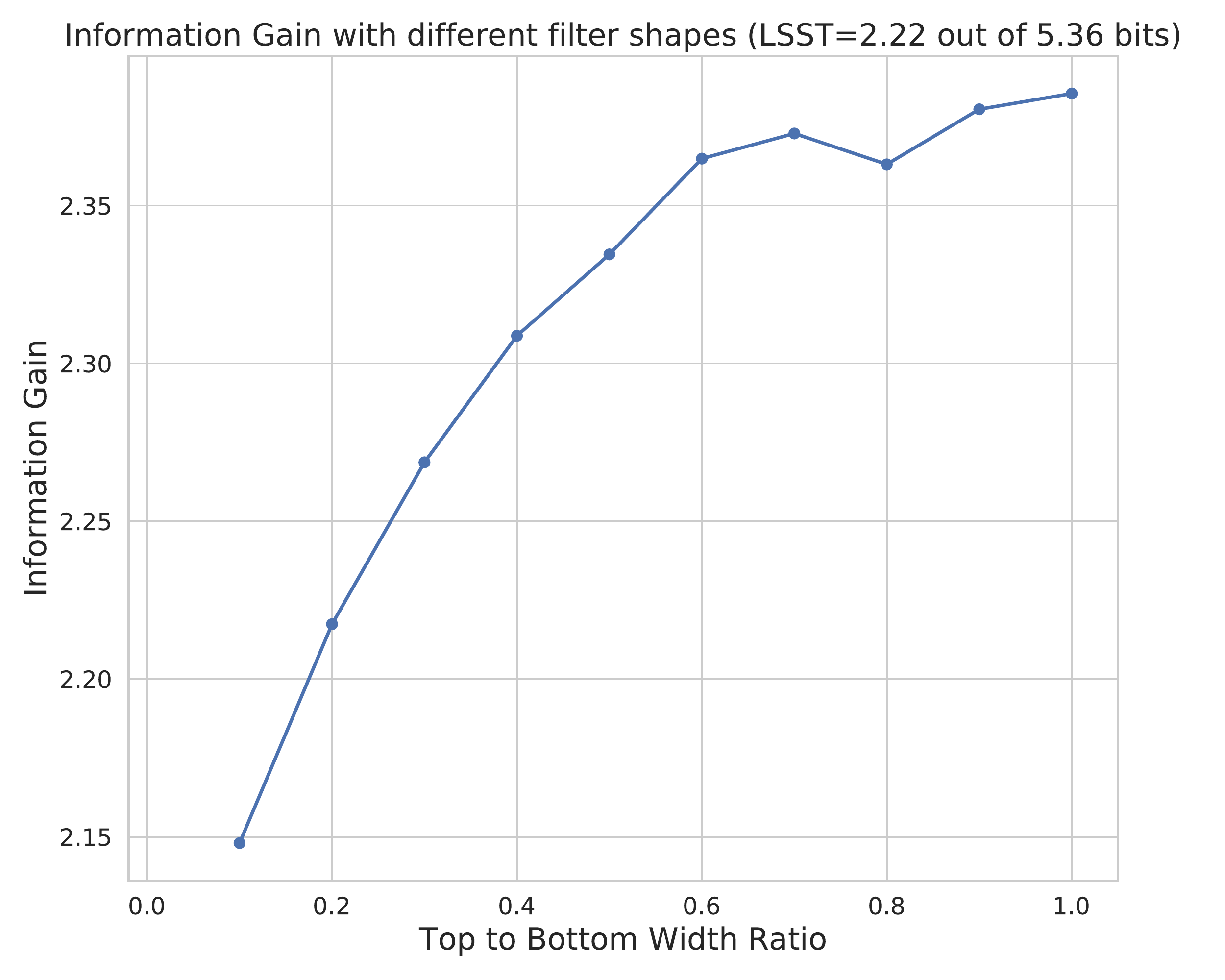} 
 \caption{The best information gain for a set of trapezoidal filters as a function of the ratio of the width for the top of the filter transmission curve to the bottom width.
   \label{fig:6_filter_results} }
\end{figure}

\begin{figure}
 \centering
 \includegraphics[width=\textwidth]{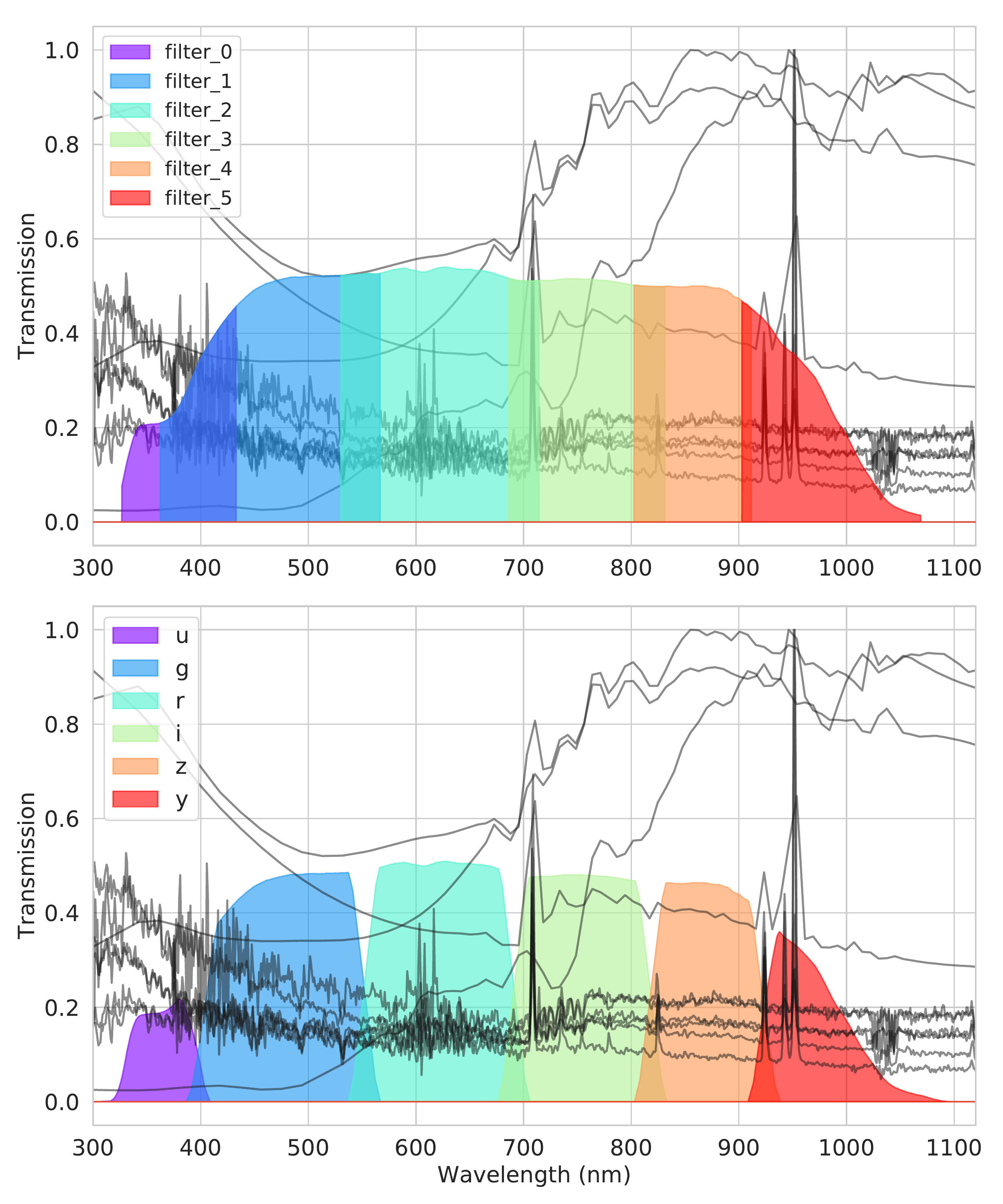} 
 \caption{Top: The best set of six filters based upon our optimization runs. These trapezoidal filters have a top-to-bottom width ratio of 1.0 and an information gain of 2.39 bits compared to the 2.22 bits of information gain from the LSST filters. Bottom: The LSST filters shown for comparison.
   \label{fig:best_new_6} }
\end{figure}

In this test we ran 10 sets of filter optimization allowing the width and locations of six filters to vary for a total of 12 degrees of freedom. In each test we set a different ratio from 0.1 - 1.0 for the top-to-bottom width of a trapezoidal filter. Figure~\ref{fig:filt_shape_compare} compares the allowable shapes for the filters with the most triangular filter having a ratio of 0.1 on the left compared to the most rectangular on the right with a ratio of top width to bottom width of 1.0.
We then found the best information gain for each filter shape at the end of the optimization run and compared the best values for each width ratio. The results are shown in Figure~\ref{fig:6_filter_results} and discussed in the sections below. Our best information gain for 6 filters was 2.39 bits which is an increase of 0.17 bits or 7\% compared to the 2.22 bits of information gain when using the LSST filters. The best performing filters are shown in the top panel of Figure~\ref{fig:best_new_6} with the LSST filters shown in the bottom panel for comparison.

\subsubsection{Filter Shape}

The results in Figure~\ref{fig:6_filter_results} clearly show a general trend that increasing the steepness of the wings of a trapezoidal filter leads to better information gain up to a ratio of 0.9 where the trend flattens out. This overall trend makes sense since the information gain is related to the width of the distributions as we saw in Figure~\ref{fig:toy_3_colors}. There, the better information gain came when the possible distribution of colors for a given redshift was narrower. This means that the width of the color distribution is affected by the signal to noise of the magnitude measurement in each filter. Allowing a wider top of the filter increases the overall transmission for filters of a similar width and thus increases the signal to noise of the flux measurement in that filter for a given spectrum at a given brightness.

Wider tops to the filters also avoids gaps in between filters without the need for a lot of overlap in the wings of each filter. Gaps in the filters allow strong features to fall between filters and wastes information that would otherwise be available. Preventing gaps is necessary to avoid this, but as we will explain in \S~\ref{subsec:filter_overlap} some overlap is beneficial but there is a limit. Narrow filter wings avoid the information loss caused by filter gaps while minimizing the extra amount of filter overlap that provides redundant information.

\subsubsection{Filter Overlap} \label{subsec:filter_overlap}

Filter sets with overlap perform better since overlapping adds information as to the position of a spectral feature in a filter. If filters do not have any overlap then it is harder to distinguish at what redshift a feature in the spectrum passes from one filter to another. In our optimal filter set every filter overlaps with its neighbors.

\begin{figure}
 \centering
 \includegraphics[width=\textwidth]{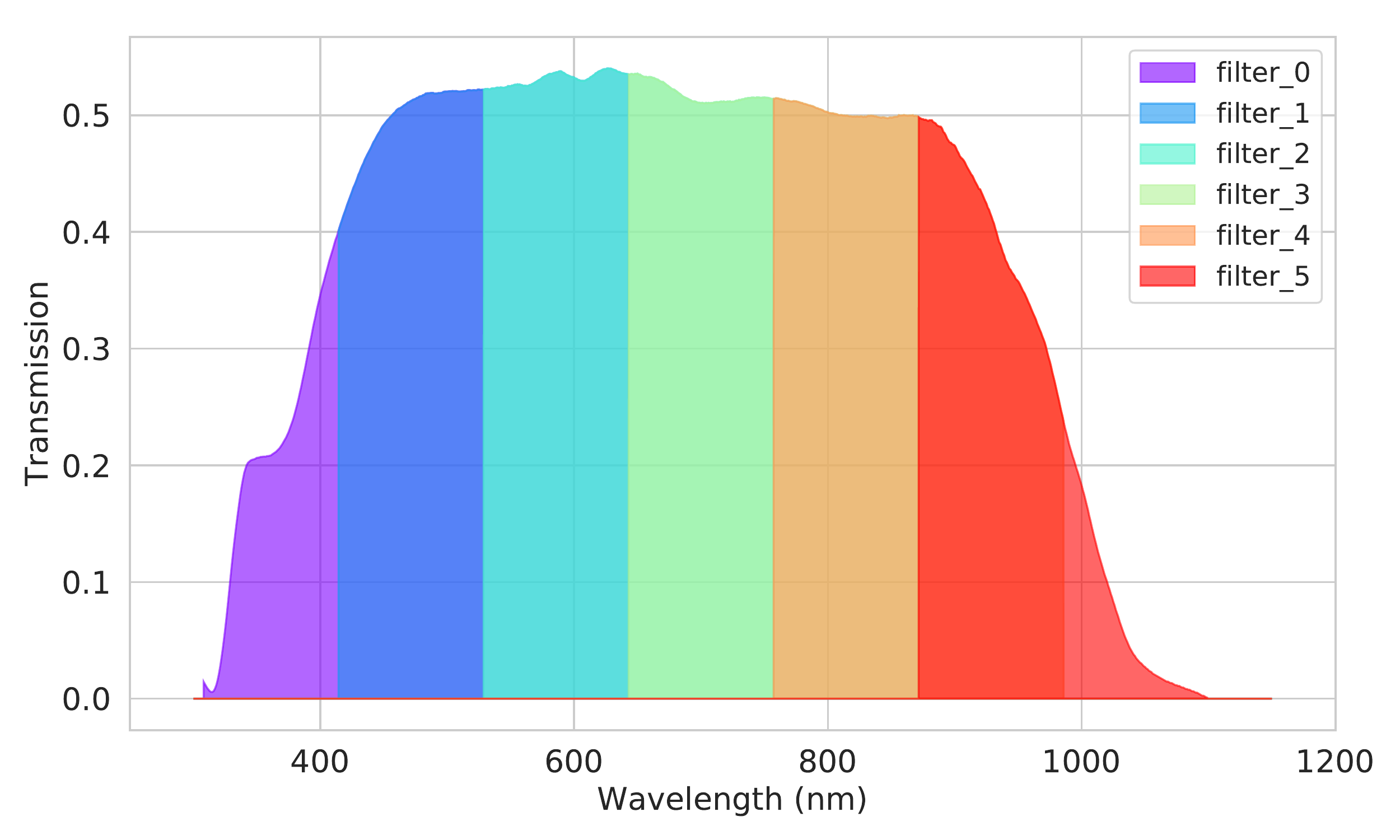} 
 \caption{6 filters with 50\% overlap of each adjacent filter. The information gain for this situation is only 1.99 bits out of 5.36 possible compared to the 2.39 bits gained with the ideal filter set that has the same top-to-bottom ratio of 1.0.
   \label{fig:complete_overlap} }
\end{figure}

However, too much overlap creates redundant information and stops being beneficial. We set up an extreme situation with a top-to-bottom ratio of 1.0 just like the optimum filters but with an overlap of half of each filter width and shown in Figure~\ref{fig:complete_overlap}. In this setup every wavelength has coverage in two filters. When we calculate the information gain for this situation it has fallen compared to the optimal filter situation above from 2.39 bits to 1.99 demonstrating that complete overlap in every possible wavelength is not ideal.

\section{Simulated photometric redshift estimation} \label{sec:simulated_photoz}

To relate the improvement in the information gain over the LSST filters to photometric redshift performance we created a simulated catalog with magnitudes measured for the LSST and new filters.

\subsection{Simulated Catalog} \label{subsec:sim_catalog}

We generated simulated catalogs of a circular area on the sky with a radius of 0.8 degrees using the LSST Catalog Simulations (CatSim) code \citep{Connolly+2014}. We generated three different catalogs, one each for the different filter sets: LSST only, 6 new filters, and LSST+1 filter designs. The LSST CatSim code generates galaxy photometry using templates from \citet{Bruzual+2003}. We ran the code with the different filter sets over the same simulated footprint over a simulated 10 year survey with the same survey properties as given in \citet{Graham+2018}. Where we included a seventh filter we gave it 160 visits to match the number in the LSST $y$ filter. Following the same procedure as \citet{Graham+2018} we included fainter galaxies and used the simulated magnitude errors to apply a random normal scatter to the catalog before making a cut at LSST $i < 25.3$. Then we made a final cut and only kept objects with a redshift $<= 2.3$ since this was the range of our redshifts when optimizing the filters in Section \ref{sec:realistic_sample}. This final cut was then split to give us 61,484 test objects and a training catalog with 301,400 objects in our simulated catalogs.

\subsection{Calculating Photometric Redshifts} \label{subsec:sim_photo_z}

We used the Color Matched Nearest Neighbors (CMNN) redshift estimation code
of \citet{Graham+2018} on our simulated data. 
The CMNN photo-z code 
calculates the Mahalanobis distance in color space between each test galaxy
and galaxies in the training catalog. The photo-z value for the test galaxy
is then the redshift of nearest neighbor in the training catalog. 
\changes{The CMNN photo-z code is designed to enable rapid characterization 
of the relative performance of photo-z estimation for catalogs with different 
filter designs (rather than the absolute photo-z accuracy). A comparison of 
the accuracy of CMNN  compared to other template fitting and machine learning
based techniques is presented in  Schmidt et al. (2019, submitted to MNRAS). 
In this paper the accuracy of CMNN, for the metrics we use in this paper,
was comparable or better than standard template based approaches.}

\begin{figure}
 \centering
 \includegraphics[width=\textwidth]{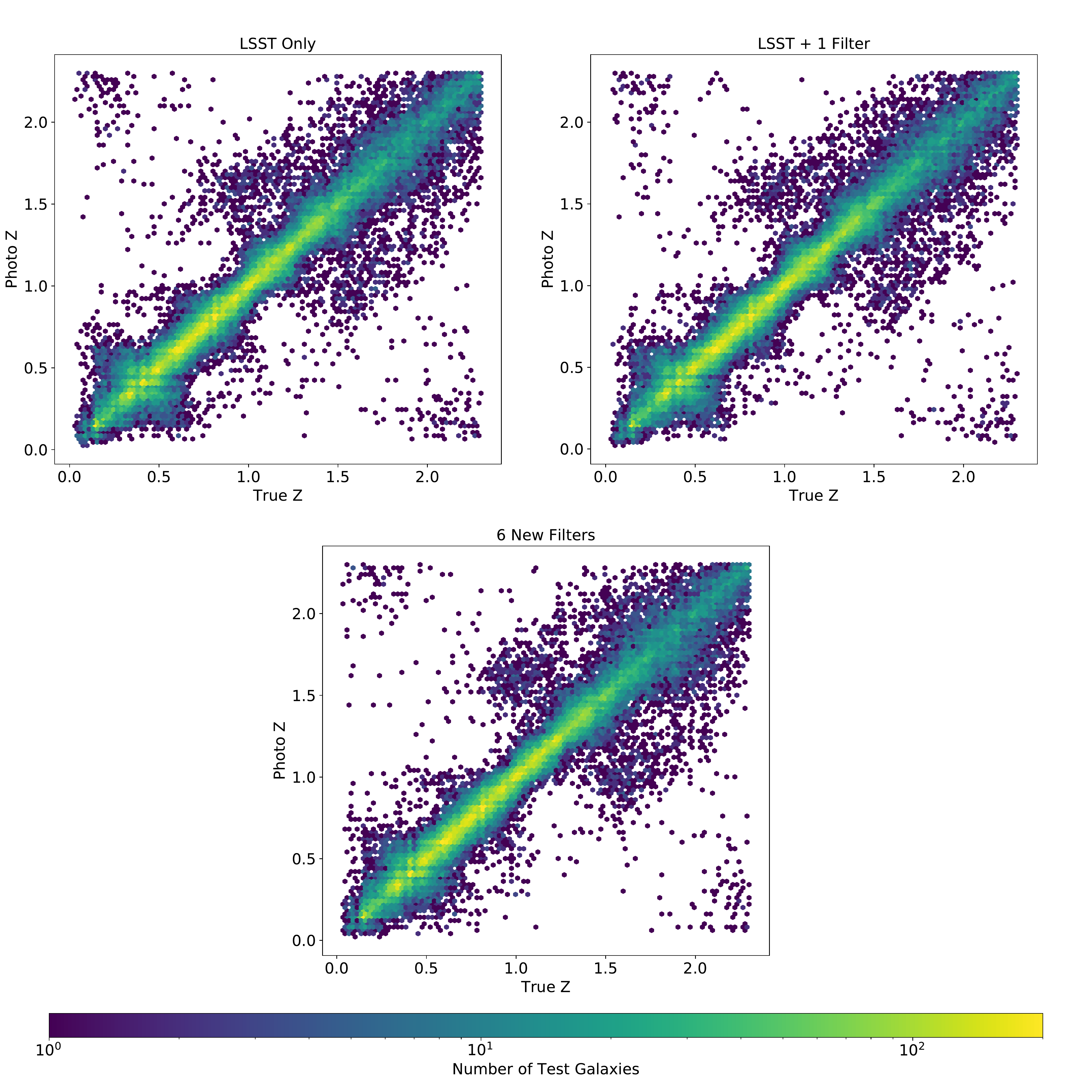} 
 \caption{Density plots for the results from photometric redshift estimation on the simulated catalogs with the CMNN photo-z code and the different filter sets. Top Left: LSST Filters Only. Top Right: LSST + 1 new filter. Bottom: 6 New Optimized Filters.
   \label{fig:pz_density_plots} }
\end{figure}

We ran the CMNN
photo-z code on each of our 3 simulated catalogs and compared the results.
Figure \ref{fig:pz_density_plots} shows the density plots comparing the input catalog
redshifts to the photometric redshifts. Between the 2 LSST based filter schemes there
does not appear to be much difference in the density plots. The 6 new filters do seem
to improve the results at redshifts $z < 0.6$
where there is a clear increase in photo-z scatter for $0.2 < z < 0.6$
visible in a cross like feature in the density plots. Beyond this the
density plots once again look similar to the LSST except there does seem to be more scatter
at redshifts greater than 1.5 where the Balmer break leaves the optical range. Since
we have previously shown that the information gain is strongly affected by the Balmer
break this is not surprising.

\begin{figure}
 \centering
 \includegraphics[width=\textwidth]{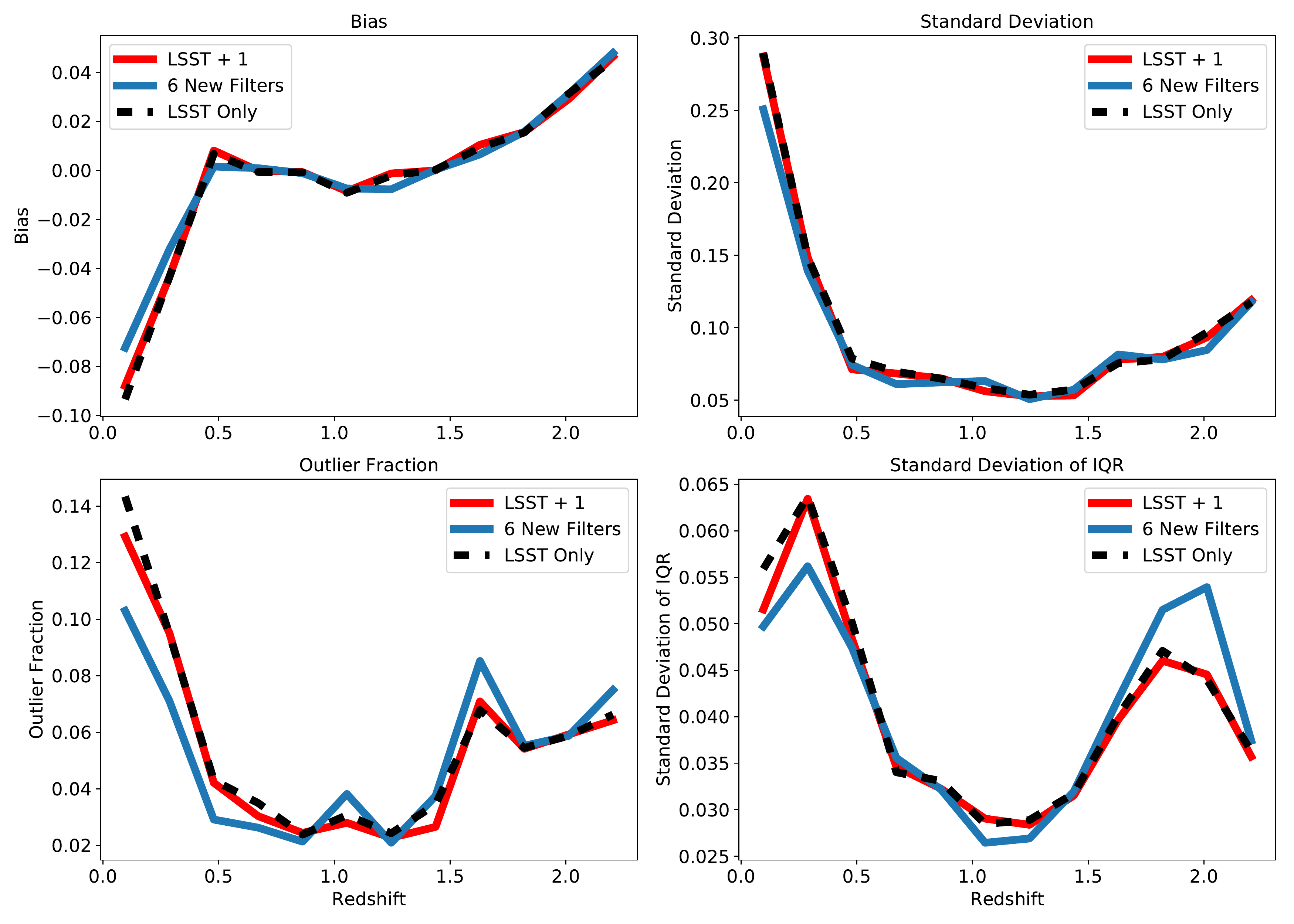} 
 \caption{Comparing the photometric redshift errors of the 3 different filter sets. As expected from the density plots adding a new filter to LSST does not change much and the 6 new filters reduce outliers at low redshift but trade this for performance at higher redshifts.
   \label{fig:pz_error_plots} }
\end{figure}

\begin{figure}
 \centering
 \includegraphics[width=\textwidth]{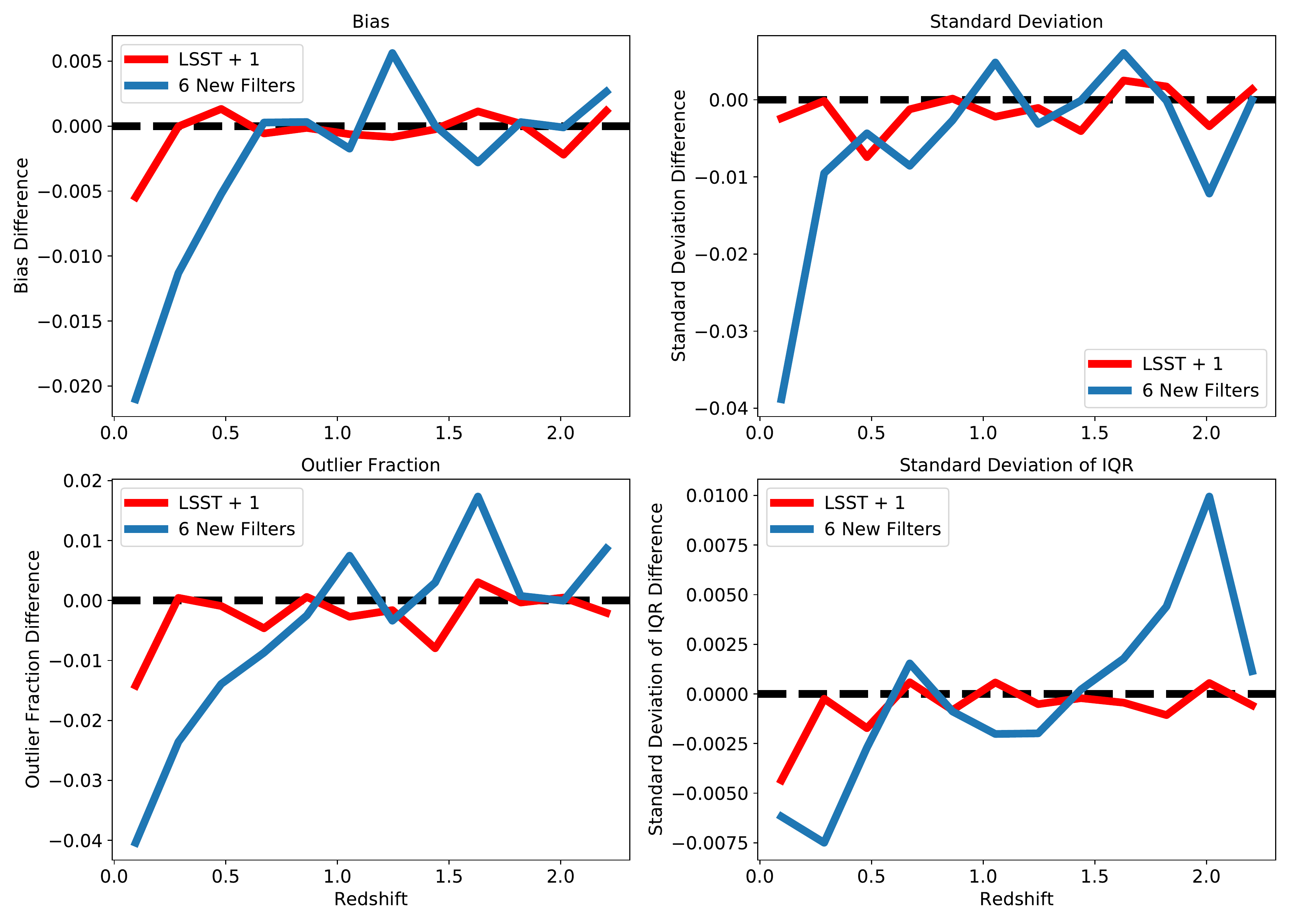} 
 \caption{Comparing the differences in photometric redshift errors of the 2 new filter sets to
 performance with the LSST filters only. The black line indicates errors are the same as the LSST filters.
 Below the black line means improvement over LSST while above the black line indicates worse performance.
 The added filter seems to slightly improve bias and overall standard deviation around the peak of the redshift prior at 0.9 and helps reduce outliers overall. As noted above the 6 new filters outperform
 LSST at lower redshifts in return for slightly degraded performance at higher redshifts.
   \label{fig:pz_error_diff} }
\end{figure}

To get a more informative
look at the errors we use the photometric redshift error defined in
\citet{Graham+2018}. The photometric redshift error is defined as $\Delta z_{1+z} =
(z_{true} - z_{phot}) / (1 + z_{phot})$ and we use this to calculate four performance
metrics. To analyze the error we plot the bias (mean $\Delta z_{1+z}$) and standard deviation of the $\Delta
z_{1+z}$ values as a function of the true redshift. We plot a robust standard deviation
which is the standard deviation of the interquartile range of the errors multiplied
by 1.349 to create a value comparable to a standard deviation. Finally we plot the fraction
of outliers with $\Delta z_{1+z}$ values greater than 0.15.
We use 12 bins in the redshift range from 0 to 2.3
and plot the values in Figure \ref{fig:pz_error_plots}. In Figure~\ref{fig:pz_error_diff} we
compare the differences to the LSST values for each new filter set. The dashed black line
is set where the values are equal to the LSST so that above this line the new filter set
is worse and below the line the new filters perform better than LSST.

The new 6 filters obtained through information gain optimization do offer more improvement
compared to adding a seventh filter which is consistent with the greater information gain improvement. 
Overall the 6 new filters improve the standard deviation by 7.1\% and the
outlier fraction by 9.9\%. These gains are driven by improvements for the redshifts below
0.9 and traded for performance losses at $z > 1.5$. This was noted in the density plots 
and is consistent with information gain focusing on the presence of the Balmer break 
in the optical range.

Adding a seventh optical filter to the LSST filters offers only slight benefits. There
is a slightly lower overall standard deviation around the $z \sim 0.9$ peak of the prior
distribution we used and over the whole test set we have a 1.6\% improvement in standard deviation.
The biggest gains are a 4.6\% improvement in overall outlier fraction and a 16.6\% improvement in the bias 
with gains appearing to be spread out through the redshift range unlike the 6 new filters.

\changes{We also performed the photo-z analysis with our alternate seventh filter 
from Section \ref{subsec:lsst_plus_one} that was optimized with brighter 
templates. When comparing the photometric redshift performance between the 
narrower and wider filters we found that: for galaxies $i<23$ the fraction of 
photo-z outliers is 4.7\% worse for the wider filter, for $i>23$ the outlier 
fraction is 2.3\% better for the wider filter. For the full population of 
galaxies the wider filter has a 2.1\% smaller fraction of photo-z outliers. These 
results are consistent with our claim in Section \ref{subsec:lsst_plus_one} that 
the width of the seventh filter is driven by the faintness of the galaxies we 
targeted with our optimization.}

\section{Discussion} \label{sec:discussion}

We were only able to provide small improvements over the LSST filters both in
terms of information gain and actual photometric redshift estimation. This is because the LSST filters
already have similar features to those we identified as optimal for photometric redshift filters. The
LSST filters have no gaps between them and each filter has a small degree of overlap with the
adjacent filters. The filters are also nearly top hat in shape with slight wings on each side which is
consistent with our findings of optimal filter sets. Our results from adding a seventh filter in the current LSST wavelength range show that another filter in optical wavelengths is not
a good way to improve photometric redshifts with LSST. In fact, it seems that since the Balmer
break is so important as we have shown that following it into the infrared regime is essential to 
improve LSST photometric redshifts. This is in line with previous work highlighting the
potential photometric redshift improvement from combining
LSST observations with infrared data from future space telescope missions
\citep{Jain+2015, Rhodes+2017, Graham+2019}.
\changes{In a future paper we will look at how much information is gained from adding infrared and UV filters
as well as what optimal filters in these wavelengths designed to complement LSST would look like.}

The practical application of the optimal filters in our work shows that our method has merit and can be 
used to design observations tailored to photometric redshift estimation. Increasing information
gain did correlate to an improvement in photo-z performance. However, limiting ourselves to 
redshifts up to $z = 2.3$ we are not able to provide insight into how filters can improve the Lyman vs 
Balmer break degeneracy. This is a large problem in photometric redshift estimation and exploring filter 
design with our method at a larger redshift range could provide interesting insights to this. Gathering 
templates that apply to the blue end of the optical range beyond $z = 2.3$ and to higher 
redshifts will help us explore the question in the future.

We used a simple redshift prior in this work,
but many options to enhance the priors we use exist. 
For example, instead of sampling each template with a uniform probability 
we could include a prior to weight certain 
templates of galaxy types more heavily at different redshifts. 
\changes{More descriptive priors reduce the starting value of entropy
for filter optimization and thus change the amount of information gain
possible from the filter design. In our catalog used in Sections
\ref{sec:realistic_sample} and \ref{sec:simulated_photoz} we had
46 redshift bins. With a flat prior where every redshift is equally
likely this equals 5.52 bits of entropy representing the maximum
level of uncertainty in the problem. With the prior based upon the
training data we reduced this to 5.36 bits of entropy. One way of looking
at this is that the prior itself provides 0.16 bits of information gain.
Then our best set of filters was able to reduce the entropy a further 2.39
bits.}
Introducing more advanced priors could 
help tailor our code to produce filters truly optimized 
for the practical application of photometric redshift estimation.
In addition, optimal photo-z filters might be different from the broadband filters we used in this work. 
We could look at how a large number of narrowband filters or a comb filter would be optimized 
for photometric redshifts. Or we could move away from trapezoidal filters and allow more 
complex shapes in the design of an individual filter.

We applied our methodology to galaxies and photometric redshifts but we can easily apply it to 
any set of templates to find the ideal filters that will differentiate between the 
corresponding astronomical objects. Future work will focus on applying the information gain 
methodology to design filters that optimize properties beyond photometric redshifts such as 
stellar observations or quasar selection. For example, templates for different stellar types 
could be used to design filters that optimize observations to determine stellar properties. 

\section{Conclusion} \label{sec:conclusion}

We have introduced a new technique to apply information theory to the design of filters in order to optimize photometric redshifts. We showed its theory and provided insight into its use with three simple examples before using it in a practical situation. We created an optimal set of six filters to cover the optical wavelengths in an ideal manner for photometric redshifts. This application revealed the general attributes of an ideal filter set for photometric redshifts. Ideal filters will have narrow wings and be near top hat in shape. They will also have a small amount of overlap. We showed that the main information for photometric redshift estimation comes from the Balmer break and optimal filters will
focus on maximizing information gain of this break at the peak of the redshift distribution.

We applied the two different filter sets to a simulated catalog of a sample of photometric data and compared the photometric redshift estimation results to the LSST filters. We showed that a set of six filters optimized using information gain could improve the standard deviation of the errors associated with photometric redshifts by 7.1\% overall at redshifts up to 2.3 over the LSST filters and outliers up to 9.9\%, but improved performance at lower redshifts was traded
for slightly worse results than LSST at higher redshifts. The LSST filters perform near the optimal set and have features we identify as optimal for photo-z filters such as overlap with neighboring filters and a nearly rectangular shape. We also discuss future directions that will improve our technique and will be possible with improvements to the code we used in this work. This python code, \textit{SIGgi}, is publicly available at \url{https://github.com/dirac-institute/siggi} and is pip installable.

This work was supported by the U.S. Department of Energy, Office of Science, under Award Number DE-SC-0011635. JBK and AJC also acknowledge support from the DIRAC Institute in the Department of Astronomy at the University of Washington. The DIRAC Institute is supported through generous gifts from the Charles and Lisa Simonyi Fund for Arts and Sciences, and the Washington Research Foundation.

\software{scikit-optimize \citep{scikit-opt}, SIGgi v0.1.6 \citep{Kalmbach2019}, Scikit-Learn \citep{scikit-learn}, LSST\ Simulations\ Software\ Stack \citep{Connolly+2014}, Numpy \citep{numpy}, Matplotlib \citep{matplotlib}, Scipy \citep{scipy}}, Pandas \citep{pandas}


\bibliography{siggi}{}

\end{document}